\documentclass[%
reprint,
 floatfix,
longbibliography,
amsmath,amssymb,
aps,
pre,
]
{revtex4-1}

\usepackage{graphicx}
\usepackage{dcolumn}
\usepackage{bm}
\usepackage[colorlinks=true, citecolor=red, linkcolor=blue,urlcolor=blue]{hyperref}
\usepackage{soul,xcolor}

\usepackage{lmodern,pdftexcmds,stmaryrd,siunitx,xcolor}
\DeclareMathAlphabet{\mathsfbi}{OT1}{\sfdefault}{bx}{sl}
\DeclareMathVersion{sfletters}
\SetSymbolFont{letters}{sfletters}{OML}{ntxsfmi}{b}{it}
\def\v{\vspace{1cm}}

\newcommand{\bcdot}{\bm{\cdot}}
\newcommand{\bnabla}{\bm{\nabla}}
\newcommand{\btimes}{\bm{\times}}

\usepackage{scalerel}
\usepackage{tikz}
\usetikzlibrary{svg.path}

\definecolor{orcidlogocol}{HTML}{A6CE39}
\tikzset{
  orcidlogo/.pic={
    \fill[orcidlogocol] svg{M256,128c0,70.7-57.3,128-128,128C57.3,256,0,198.7,0,128C0,57.3,57.3,0,128,0C198.7,0,256,57.3,256,128z};
    \fill[white] svg{M86.3,186.2H70.9V79.1h15.4v48.4V186.2z}
                 svg{M108.9,79.1h41.6c39.6,0,57,28.3,57,53.6c0,27.5-21.5,53.6-56.8,53.6h-41.8V79.1z M124.3,172.4h24.5c34.9,0,42.9-26.5,42.9-39.7c0-21.5-13.7-39.7-43.7-39.7h-23.7V172.4z}
                 svg{M88.7,56.8c0,5.5-4.5,10.1-10.1,10.1c-5.6,0-10.1-4.6-10.1-10.1c0-5.6,4.5-10.1,10.1-10.1C84.2,46.7,88.7,51.3,88.7,56.8z};
  }
}

\newcommand\orcidicon[1]{\href{https://orcid.org/#1}{\mbox{\scalerel*{
\begin{tikzpicture}[yscale=-1,transform shape]
\pic{orcidlogo};
\end{tikzpicture}
}{|}}}}

\usepackage{hyperref} 

\begin{document}
\def\v{\vspace{1.5cm}}

\setstcolor{red}


\title{Transition to bound states for bacteria swimming near surfaces}

\author{Debasish Das~\orcidicon{0000-0003-2365-4720}}
\email{dd496@damtp.cam.ac.uk}
\author{Eric Lauga~\orcidicon{0000-0002-8916-2545}}%
 \email{e.lauga@damtp.cam.ac.uk}
\affiliation{%
Department of Applied Mathematics and Theoretical Physics, Centre for Mathematical Sciences, University of Cambridge, Wilberforce Road, Cambridge CB3 0WA, UK.}%

\date{\today} 

\begin{abstract} 
It is well known that flagellated bacteria swim in circles near surfaces. However, recent experiments have shown that a sulfide-oxidizing bacterium named \textit{Thiovulum majus} can transition from swimming in circles to a surface bound state where it stops swimming while remaining free to move laterally along the surface. In this bound state, the cell rotates perpendicular to the surface with its flagella pointing away from it. Using numerical simulations and theoretical analysis, we demonstrate the existence of a fluid-structure interaction instability that causes cells with relatively short flagella to become surface bound.  
\end{abstract}

\pacs{Valid PACS appear here}
\maketitle

\section{Introduction}\label{sec:intro}

Bacteria constitute the larger of the two domains of prokaryotic organisms---unicellular organisms lacking a distinct nucleus and other membrane-bound organelles. Being one of the first organisms to have appeared on Earth, they have evolved to thrive in a variety of  environments. Their shapes can range from rods and spirals to spheres, depending on their surroundings and ecological niches. Though bacteria differ greatly in size, the majority measure only a few micrometers and chiefly rely on diffusion to transport metabolites. While the smallest bacteria, \textit{Mycoplasma genitalium}, range in size from 0.2--\SI{0.3}{\micro\meter}, a few other bacteria have evolved to a rather unusual gigantic size \cite{schulz2001}, sometimes reaching up to \SI{750}{\micro\meter} \cite{schulz1999}. Such exotic bacteria have received far less attention when compared to the more commonly encountered species such as \textit{Escherichia coli} \cite{berg2008}.  Bacterial gigantism can be advantageous for enhanced resistance to predation but it comes at the price of reduced nutrient uptake  due to a lower cellular surface area to volume ratio and the need for  different strategies to cope with it.

\begin{figure}[b]
    \centering
    \includegraphics[width=0.95\linewidth]{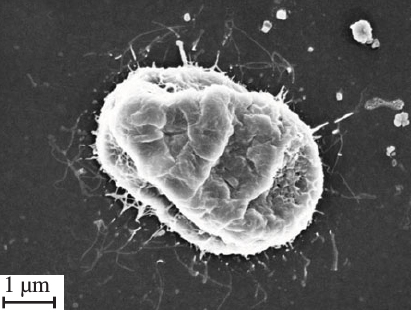}
    \caption{Scanning electron microscope  image of a \textit{Thiovulum majus} bacterium with a slightly shrunken spherical cell body of radius $a\sim$ \SI{4}{\micro\meter} and flagella $L \sim$ \SI{2}{\micro\meter} long, reproduced with permission from Ref.~\cite{petroff2015a} (see similar  pictures in Ref.~\cite{deboer1961}).}
    \label{fig:experimentalphoto}
\end{figure}

In this article, we focus on the locomotion of one such outlier species of bacteria named \textit{Thiovulum majus} (\textit{T.~majus}),  
a sulfide-oxidizing organism generally found near hydrogen sulfide deposits in seas or marshes~\cite{hinze1913,wirsen1978,garcia1989}. It is nearly spherical in shape and has a reasonably large cell body, typically between 5--\SI{25}{\micro\meter} (Fig.~\ref{fig:experimentalphoto}). More importantly, it is the second fastest swimming bacterium in nature with speeds reaching up to \SI{615}{\micro\meter\per\second} \cite{fenchel2004}. Thus, \textit{T.~majus} cells are able to overcome the limitations in nutrient diffusion by actively stirring up the fluid medium. When attached to surfaces via mucus threads, the cells continue rotating their cell body and flagella to generate advective oxygen transport   about 40 times higher than that generated by molecular diffusion, thus  significantly enhancing their nutrient uptake \cite{fenchel1998}. 
 

In homogeneous oxygen concentrations, \textit{T.~majus} swim in helical trajectories and rarely change their direction \cite{fenchel1994}, quite unlike the run-and-tumble motion of \textit{E.~coli}. In recent experiments, the swimming behavior of \textit{T.~majus} cells  was studied near surfaces \cite{petroff2015b,petroff2018} (henceforth, we will use the words surface and wall interchangeably). Quite surprisingly, it was found that instead of swimming along the surface in circles, some freely swimming cells became dynamically surface bound. In this bound state, cells remained free to move laterally along the surface and their bodies continuously rotated around their center in the direction perpendicular to the surface while their flagellar filaments pointed away from the surface, rotating in the opposite direction.

\begin{figure}[t]
    \centering
    \includegraphics[width=0.95\linewidth]{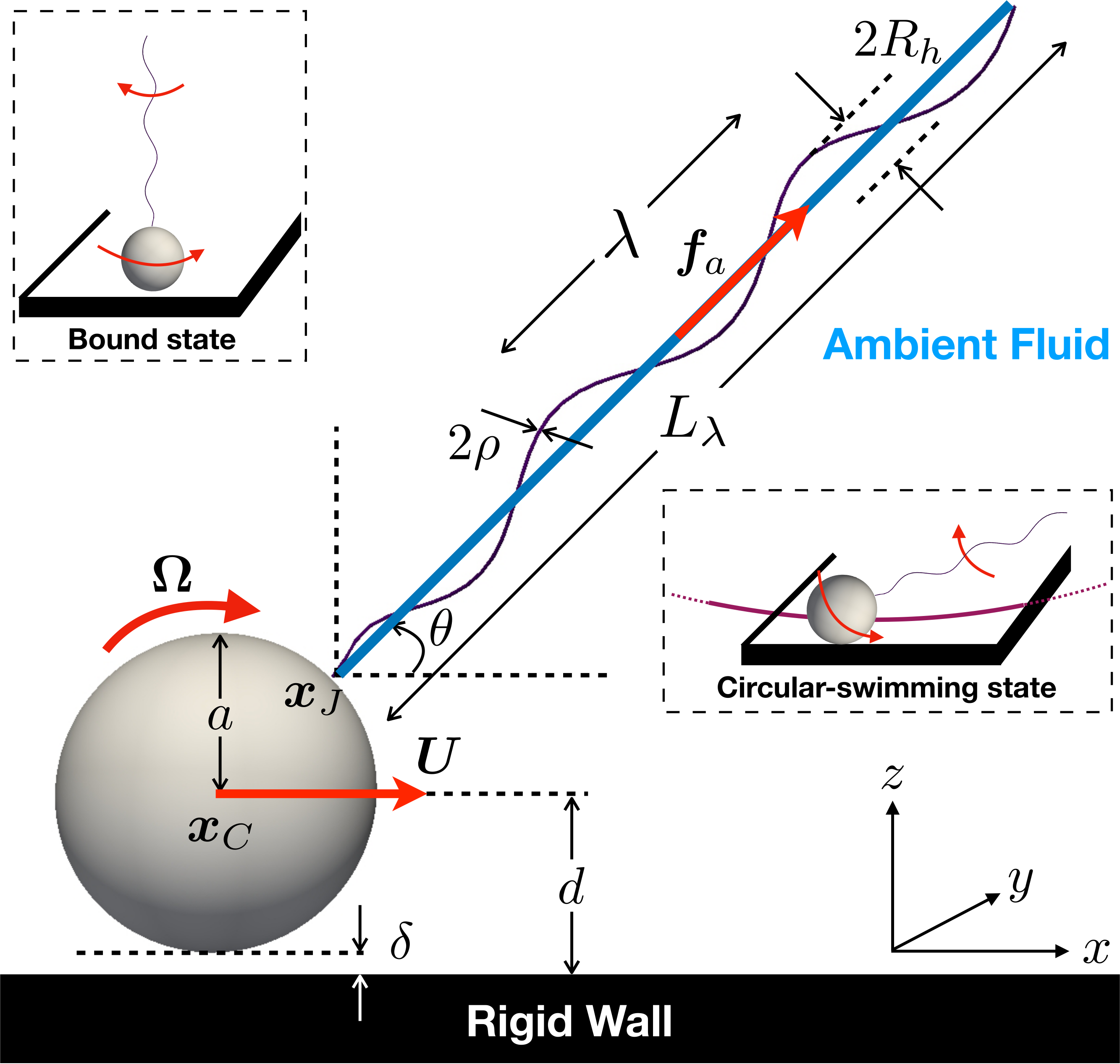}
    \caption{Schematic diagram of a  flagellated bacterium swimming near a plane surface. The cell body is modeled as a rigid sphere and the flagellar filament as either a rigid right-handed helix rotating along its axis for the numerical simulations in \S\ref{sec:numericalmodel} or a rod placed along the helix axis with an active force $\bm{f}_a$ acting along it for the theoretical model in \S\ref{sec:theorymodel}. The cell body undergoes rigid body translation and rotation with velocities $\bm{U}$ and $\bm{\Omega}$, respectively. The flagellum tilt angle is denoted by $\theta$ (see \S \ref{sec:numericalmodel} for description of other notations). The two possible final steady states of a bacterium near a surface are shown in the insets, namely, the surface-bound and circular-swimming states.}
    \label{fig:schematic}
\end{figure}

The bound state is in stark contrast to bacteria swimming in circular paths near surfaces~\cite{berg1990,ramia1993,lauga2006}. The question therefore arises regarding the mechanism at the origin of this transition to a bound state.  Mathematically, we define the cell to be in the surface-bound state when the flagellum axis is perpendicular to the surface, i.e.,~$\theta=\pi/2$ (Fig.~\ref{fig:schematic}), and consequently, the radius of circular trajectory is $R=0$. In the bound state, a small perturbation in the tilt angle of the flagellum is expected to destabilize the cell and cause it to swim parallel to the surface in circles. So what makes this state stable? Using a combination of theory and simulations, we show that the transition from swimming to a bound state can be rationalized as an instability due to fluid-structure interaction.

Specifically, we show that the flagellum of a freely swimming \textit{T.~majus} cell undergoes slow tilt angular dynamics near a rigid wall. If the distance between the cell's surface and the wall becomes sufficiently small, drag forces acting on the translating flagellum (which tend to align the flagellum parallel to the surface) are unable to compensate the large lubrication torque exerted on the cell body (which tends to align the flagellum perpendicular to the surface). In this case, the cell eventually points perpendicular to the wall and thus stops swimming, while the cell body and flagella continue rotating as seen in experiments~\cite{petroff2015b,petroff2018}. This bound state is stable only below a certain critical flagellum axial length, $L_\lambda$, normalized by the radius of the spherical cell body, $a$, consistent with experiments, since \textit{T.~majus} cells have relatively short flagella and large spherical cell bodies, $L_\lambda/a=0.5$ (Figs.~\ref{fig:experimentalphoto} and Fig.~\ref{fig:schematic}).

The article is organized as follows. We first define the problem in \S \ref{sec:problem}, then describe the numerical and theoretical models in \S \ref{sec:numericalmodel} and \S \ref{sec:theorymodel}, respectively. Next, we present the results in \S \ref{sec:results} and finally conclude in \S \ref{sec:conclusion}.

\section{Problem setup}\label{sec:problem}
The fluid dynamics of bacteria, owing to their relatively small size, 
are described by the incompressible Stokes equations, $-\bnabla p + \mu \nabla^2 \bm{u} = 0,~ \bnabla \bcdot \bm{u} = 0$, where $p$ and $\bm{u}$ are the dynamic pressure and velocity of the fluid \cite{lauga2009,lauga2016}. The surface of a \textit{T.~majus} cell is covered with about 100 flagella, but neither their precise number  nor their rotation rates have been measured. 
However, these spherical cells rotate in a counterclockwise direction when viewed from the posterior side \cite{fenchel2004,petroff2015b}, which  motivates us to model the fraction of flagella on the cell surface that cause propulsion bundled together as a single right-handed clockwise rotating helix for the numerics in \S \ref{sec:numericalmodel} or rigid active pushing rod for the theory in \S \ref{sec:theorymodel}. The cell body is assumed to be a sphere of radius $a$, also chosen as the length scale for the problem. The sphere is centered at $\bm{x}_C$ while the flagellum is attached to the cell body surface at the junction point, $\bm{x}_J$ (Fig. \ref{fig:schematic}). The minimum distance between the spherical surface and the plane rigid wall is $\delta$ and the distance between the center of the sphere and the wall is $d=a+\delta$. The tilt angle of the flagellum measured with respect to the horizontal direction is denoted by $\theta$, so that $\theta=\pi/2$ and $\theta\neq\pi/2$ correspond to surface-bound and circular-swimming states, respectively.

\section{Numerical Model}\label{sec:numericalmodel}

The flagellum is modeled as a rigid right-handed helix with tapered ends \cite{higdon1979a,higdon1979b},
\begin{equation}
\bm{x} = [l,E(l)R_h\cos(kl+\psi),E(l)R_h \sin(kl+\psi)],
\end{equation}
where $l \in [0,L_\lambda]$ and $k$ is the helix wavenumber (Fig.~\ref{fig:schematic}). The tapering function is $E(l)=1-\mathrm{e}^{-k_E^2 l^2}$, where $k_E=k$ is a constant determining how quickly the helix grows to its maximum amplitude, $R_h$. The total axial and contour length of the helix are $L_\lambda$ and $L=L_\lambda/\cos\phi$, respectively, where $\phi= \arctan{(R_hk)}$ is the pitch angle. Changing the phase angle, $\psi$, simply rotates the helix around its axis. The axial wavelength and wave speed are $\lambda = 2\pi/k$ and $V=\omega/k$, respectively. The cross-sectional radius and aspect ratio of the helix are $\rho$ and $\varepsilon = \rho/L$, respectively. 

The hydrodynamics of the effective propelling flagellum are described using 
slender-body theory \cite{johnson1980}. Under this framework, the velocity $\bm{u}$ of the centerline $C_f$ of a slender helix is related to the hydrodynamic force per unit length, $\bm{f}_h$, acting on it through the integral equation,
\begin{equation}\label{eq:SBTequation}
\begin{split}
\bm{u}(\bm{x}_0) = -\frac{1}{8\pi \mu}\bm{\Lambda}[\bm{f}_h](\bm{x}_0)- \frac{1}{8\pi \mu}\bm{K}[\bm{f}_h](\bm{x}_0),
\end{split}
\end{equation}
where $\bm{x}_0 =\bm{x}_0(s,t)$ and $s \in [0,L]$. 
The no-slip  boundary condition on the rigid wall is satisfied by using appropriate image singularities \cite{blake1974}. The exact expressions of $\bm{\Lambda}$ (local operator) and $\bm{K}$ (non-local operator) are provided in Appendix \ref{sec:sbtappendix}.


In recent modeling work involving locomotion of bacteria and spermatozoa, boundary integral equations for the cell body combined with slender-body theory for the flagellum have proved to be accurate and efficient \cite{das2018,smith2009}. However, when resolving cell-wall interactions, extremely fine surface grid resolution becomes necessary when the cell body gets very close to the wall, $\delta^* \sim 0.0001$ (see discussion in Appendix \ref{sec:bemtest}). Hence, dynamical simulations involving iterations in time become prohibitively expensive. In order to circumvent this issue, the hydrodynamics of the spherical cell body and its interaction with the rigid wall are described using analytical results in the far-field and lubrication limits. We switch between the two at a cut-off distance using exact solutions based on bispherical coordinates, see Appendix \ref{sec:rmmatrices}. The hydrodynamic interactions between the cell body and flagellum are unaccounted for but are not expected to alter the physics of the problem. 
The force and torque on a sphere are related to the translational and angular velocities about the junction $\bm{x}_J$ by a symmetric resistance matrix
\begin{equation}
\begin{bmatrix} 
\bm{F} \\ 
\bm{T}
\end{bmatrix}
 =
  \begin{bmatrix}
   \mathsfbi{A} & \mathsfbi{B} \\
   \mathsfbi{B}^T & \mathsfbi{C} \\
  \end{bmatrix}
  \begin{bmatrix} 
\bm{U} \\ 
\bm{\Omega} 
\end{bmatrix},
\end{equation}
with all matrix elements provided in Appendix \ref{sec:bemtest}. The kinematic boundary condition for the flagella is
\begin{equation}
\bm{u}(\bm{x}_0) = \bm{U} + (\bm{\Omega} + {\Omega}_f \bm{e}_f) \btimes (\bm{x}_0 - \bm{x}_J),
\end{equation}
where $\bm{U}$ and $\bm{\Omega}$ are the translational and angular velocities of the cell body, $\Omega_f$ is the angular velocity of the flagellar filament relative to the cell body, and $\bm{e}_f$ is the helical axis direction pointing away from the cell body. We prescribe the dimensionless value $\Omega_f=-1$ that acts as forcing for the system (a right-handed helical flagellum must rotate in a clockwise direction when viewed from the posterior end for the bacterium to be a pusher). The force and torque balance equations for the whole bacterium computed about the junction point $\bm{x}_J$ are
\begin{align}
&\bm{F} + \iint\displaylimits_{C_f} \bm{f}_h(\bm{x}_0) \,\mathrm{d}s = 0, \label{eq:forcebalance}\\ 
&\bm{T} + \iint\displaylimits_{C_f} (\bm{x}_0-\bm{x}_J) \btimes \bm{f}_h(\bm{x}_0) \,\mathrm{d}s = 0. \label{eq:torquebalance}
\end{align}
Solving Eqs.~\ref{eq:SBTequation}, \ref{eq:forcebalance} and \ref{eq:torquebalance}  numerically provides us the desired velocities, $\bm{U}$ and $\bm{\Omega}$. When $\delta^*\le 0.0001$, the validity of Stokes equations becomes questionable at such small scales. To prevent the cell from getting any closer to the wall, we add a contact force, acting on it perpendicular to the wall and passing through the cell body center, to the force and torque balance equations such that $\bm{U}\bcdot\bm{e}_z=0$. Note that there are no direct  experimental measurements of the gap thickness, however, it has been estimated to range from \SI{4.25}{\nano\meter} to  \SI{140}{\nano\meter} based on scaling arguments~\cite{petroff2018}. Once the velocities of the body are found, the center of the cell body  and the position of the flagellum  are advanced in time using a second-order Runge-Kutta time marching scheme until a dynamical steady state is reached.

\section{Theoretical Model}\label{sec:theorymodel}

The numerical model can be further simplified to elucidate analytically the fundamental mechanisms at play behind the stability of the bound states. Here, the cell body is modeled as a sphere as in \S \ref{sec:numericalmodel} while the flagellum is modeled as a rigid rod with a constant `active force', $\bm{f}_a$, acting on it (Fig.~\ref{fig:schematic}). This force represents the hydrodynamic resistive drag arising from the rotation of the helical flagellum. Hydrodynamic interactions between the active rod and the rigid wall are neglected. The cell-flagellum junction is located at $\bm{x}_J= a (\cos \theta \bm{\hat{e}}_x + \sin \theta \bm{\hat{e}}_z)$ at any given time. The dimensionless gap thickness is set to $\delta^*=0.0001$ and the bacterium is restricted to have a translational velocity in the $x$ direction only. The position and instantaneous velocity of points along the active rod are
\begin{align}
\bm{x}_f &= \bm{x}_C + (a+l)(\cos \theta \bm{\hat{e}}_x + \sin \theta \bm{\hat{e}}_z), \\
\bm{u}_f &= (U + \Omega (a+l)\sin \theta) \bm{\hat{e}}_x - \Omega (a+l) \cos \theta \bm{\hat{e}}_z,
\end{align} 
respectively, where $l \in [0,L_\lambda]$. Also, note that according to our definition $\dot{\theta} = -\Omega$. The unit  tangent and normal to the rod are $\bm{\hat{t}} = \cos \theta \bm{\hat{e}}_x + \sin \theta \bm{\hat{e}}_z$ and $\bm{\hat{n}} = -\sin \theta \bm{\hat{e}}_x + \cos \theta \bm{\hat{e}}_z$, respectively. 
\begin{figure*}
    \centering
    \includegraphics[width=0.95\linewidth]{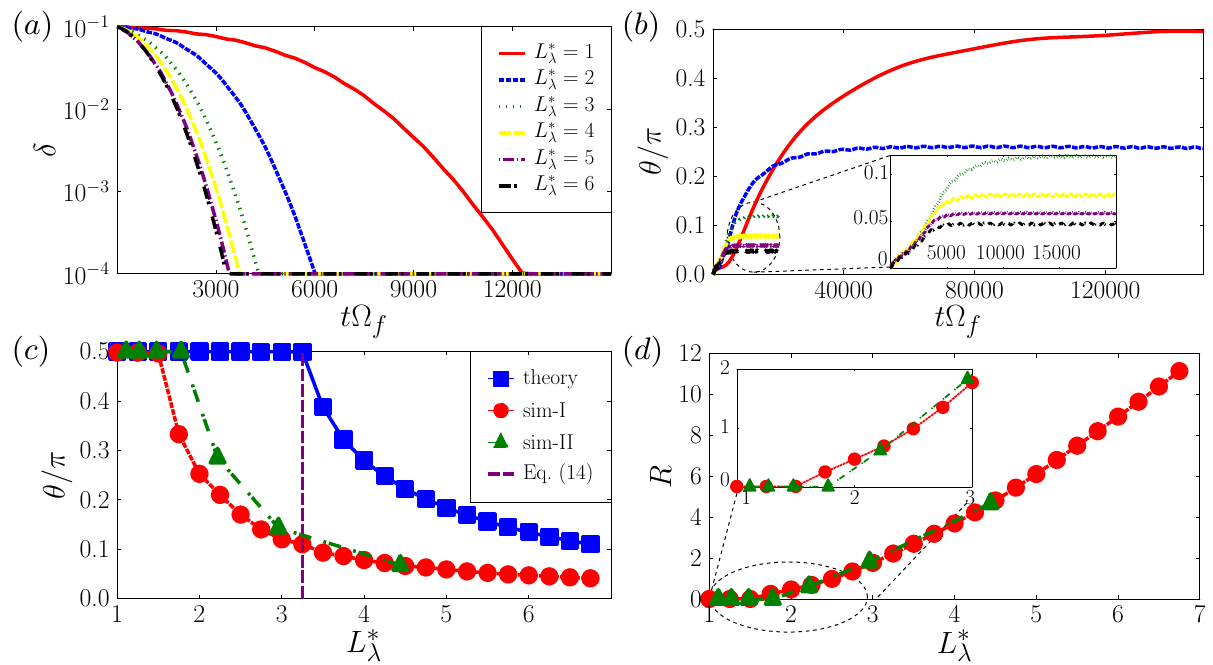}
    \caption{Temporal evolution plot of ($a$) the minimum distance between the spherical cell body surface and the rigid wall $\delta^*$, showing attraction of the bacterium towards the wall for various flagellum axial lengths,  $L_\lambda^*=1-6$ (time $t$ is scaled using $\Omega_f^{-1}$), and ($b$) the tilt angle $\theta/\pi$. ($c$) Pitchfork bifurcation of the tilt angle, $\theta/\pi$, plotted against the flagellum axial length, $L_\lambda^*$, for the numerical simulations (red dashed lines/circles for sim-I and green dash-dotted lines/triangles for sim-II) and the theoretical model (blue solid lines/squares). The critical flagellum axial length below which bound state is possible, obtained by Eq.~\eqref{eq:critical}, is shown in purple dashed line. Bound and swimming states are represented by $\theta=\pi/2$ and $\theta\neq\pi/2$, respectively. ($d$) Pitchfork bifurcation of the circular radius, $R$, traced by the bacteria when it swims parallel to the wall (numerical simulations).}
    \label{fig:results}
\end{figure*}
The parallel and perpendicular components of the drag force $d\bm{f}=d\bm{f}_n + d\bm{f}_t$ acting on the small element are
\begin{equation}
d\bm{f}_t = -c_t (\bm{u}_f\bcdot\bm{\hat{t}}) \bm{\hat{t}} dl, \quad d\bm{f}_n = -c_n (\bm{u}_f\bcdot\bm{\hat{n}}) \bm{\hat{n}} dl,
\end{equation}
which can be decomposed into directions parallel and perpendicular to the wall, $d\bm{f}= df_x \bm{\hat{e}}_x + df_z \bm{\hat{e}}_z$.
The net force and net torque acting on the rod in the $x$ and $y$ directions are,
\begin{equation}
F_f = \int_0^{L_\lambda} df_x, \quad T_f = \int_0^{L_\lambda} [(\bm{x}_{f}-\bm{x}_{C}) \btimes d\bm{f}_n] \bcdot \bm{\hat{e}}_y,
\end{equation}
respectively. 
The mobility matrix of a sphere next to a wall about $\bm{x}_C$ is written as
\begin{equation}\label{eq:RM2D}
\begin{pmatrix}
  aF_x  \\
  T_y  \\  
 \end{pmatrix}= -
 \begin{pmatrix}
  6\pi\mu a^3 F_{t}^* & 6\pi\mu a^3 F_{r}^*\\
  8\pi\mu a^3 T_{t}^* &   8\pi\mu a^3 T_{r}^* \\
 \end{pmatrix}
 \begin{pmatrix}
  U/a \\
  \Omega \\
 \end{pmatrix}.
\end{equation}
The expressions for the drag coefficients $c_n$ and $c_t$ and matrix elements are provided in Appendix \ref{sec:resistancerod}. The only relevant equations to capture the tilt dynamics of the cell are the total force and torque balance equations in the $x$ and $y$ directions, 
\begin{equation}
\begin{split}\label{eq:theoryforce}
& -U[6\pi\mu  F_{t}^* + (c_n \sin^2\theta + c_t \cos^2\theta)L_\lambda^*] \\
& -\Omega[6\pi\mu F_{r}^* + c_n  (L_\lambda^* + \tfrac{1}{2}L_\lambda^{*2}) \sin\theta]  + f_a L_\lambda^* \cos \theta = 0,
\end{split}
\end{equation}
and,
\begin{equation}
\begin{split}\label{eq:theorytorque}
& - U[ 8\pi\mu T_{t}^* + c_n \sin\theta (L_\lambda^* + \tfrac{1}{2}L^{*2})] \\
& -\Omega[ 8\pi\mu T_{r}^* + c_n (L_\lambda^* + L_\lambda^{*2} + \tfrac{1}{3}L_\lambda^{*3})] = 0,
\end{split}
\end{equation}
respectively, where $L_\lambda^* = L_\lambda/a$. Note, that the forcing for the system is the dimensionless active force strength, $f_a=-1$, for pushers. We can solve the coupled Eqs.~\eqref{eq:theoryforce} and \eqref{eq:theorytorque} simultaneously to obtain the body velocities, $U$ and $\Omega$, and the tilt angle $\theta$ as a function of time   until a steady state is reached. 

\section{Results}\label{sec:results}
The results of the numerical and theoretical models are illustrated in Fig.~\ref{fig:results}. 

\subsection{Numerical model}
In the numerical simulations, the cell is allowed to swim in three dimensions. The cell body is placed at a distance $d^*=1.1$ from the wall ($\delta^*=0.1$) with its flagellar filament initially parallel to the wall ($\theta=0$). Owing to a lack of experimental data, the geometrical parameters of the flagellar filament are assumed to be similar to that of \textit{E.~coli}, namely, $\rho= \SI{0.012}{\micro\meter}$, $\lambda= \SI{2.22}{\micro\meter}$, and $R_h= \SI{0.2}{\micro\meter}$ \cite{deboer1961}. In the first set of simulations (sim-I), the axial length of the filament  is varied from $L_\lambda = \SI{1.0}{\micro\meter}$ to \SI{6.75}{\micro\meter} while the cell body radius is fixed at $a= \SI{1}{\micro\meter}$. In the second set of simulations (sim-II), we fix the number of turns at $N=2$ and axial length at $L_\lambda= \SI{4.44}{\micro\meter}$ while the cell body radius is varied from $a=\SI{1}{\micro\meter}$ to \SI{4}{\micro\meter}.

The simulations are performed as described in \S \ref{sec:numericalmodel} until a dynamical steady state is reached. 
We plot in  Fig.~\ref{fig:results}(a) the dynamics of $\delta(t)$ for six values of the flagellum length, $L_\lambda^*$, for sim-I. In each case, the cell is attracted to the wall regardless of the flagellar filament length.  As the cell gets closer to the wall the coupling between translation and rotation becomes stronger which causes the filament to tilt away from the wall. This is illustrated in  Fig.~\ref{fig:results}(b) where we plot the tilt angle, $\theta(t)$, for the same six cases as in Fig.~\ref{fig:results}(a). This tendency to tilt away from the wall due to cell body-wall interaction is resisted by the viscous torque experienced by the translating flagella as well as an attractive torque (see Appendix \ref{sec:torquehelix}) that tends to align the helix parallel to the wall arising from helix-wall hydrodynamic interactions. As a result, the tilt angle reaches a dynamical steady state whose value depends on $L_\lambda^*$.

The steady state tilt angles, $\theta$, and the radius of circular trajectories for the bacterium, $R$, are shown in Figs.~\ref{fig:results}(c) and \ref{fig:results}(d), respectively, as a function of the dimensionless flagellar filament length, $L_\lambda^*$. In particular, we find that cells with axial length $L_\lambda^*\leq 1.5$ for sim-I and 1.78 for sim-II tilt up vertically and become bound to the surface ($\theta=\pi/2$ and $R=0$) while cells with longer flagellar filaments continue swimming in circular trajectories. Notably,  $\theta$ and $R$ appear to undergo a pitchfork bifurcation at these critical values. 

\subsection{Theoretical model}

The results from the theoretical model, where motion is confined to one dimension along the surface,  are shown in Fig.~\ref{fig:results}(c) with a distance between  the cell-surface and the wall   kept fixed at $\delta^*=0.0001$. We only use lubrication limit solutions for the resistance matrix in Eq.~\ref{eq:RM2D}. The tilt angle, $\theta$, plotted against $L_\lambda^*$ 
is also seen to undergo a pitchfork bifurcation from a bound to a swimming state. The critical flagellar length for bound states predicted by the theoretical model is higher than the numerical model. This is primarily due to an attractive torque arising from helix-wall hydrodynamic interaction (see Appendix \ref{sec:torquehelix}) that tends to decrease the tilt angle of the helix present in the numerical model but absent in the theoretical one. Nonetheless, 
the theoretical model is able to capture the main physics of the transition to a bound state. 

\subsection{Stability}

We may also use the theoretical model to gain insight into the stability of the cells swimming near surfaces. To do so, we linearize Eqs.~\eqref{eq:theoryforce} and \eqref{eq:theorytorque}, and perform a stability analysis around the bound state with the cell oriented perpendicular to the surface. Writing $\theta=\pi/2 + \alpha$, where $\alpha$ is a small perturbation in tilt angle, we obtain the linear dynamical equation
\begin{widetext}
\begin{align}\label{eq:critical}
    \frac{d \alpha}{dt} & = -f_a L_\lambda^*\left[ \frac{8\pi\mu T_{t}^* + c_n (L_\lambda^* + \tfrac{1}{2}L_\lambda^{*2})}{8\pi\mu T_{r}^* + c_n (L_\lambda^* + L_\lambda^{*2} + \tfrac{1}{3}L_\lambda^{*3})(6\pi\mu  F_{t}^* + c_nL_\lambda^*) - \{ 8\pi\mu T_{t}^* + c_n (L_\lambda^* + \tfrac{1}{2}L_\lambda^{*2})\}^2}\right]\alpha.
\end{align}
\end{widetext}
Since $f_a<0$ for pushers, the critical flagella length is simply found when the value of the growth term in square brackets in Eq.~\eqref{eq:critical}, denoted $\sigma$,  changes sign. We find that $\sigma<0$ for   shorter flagella and hence the system is stable in the bound state, and a transition to $\sigma>0$ occurs at a critical value of $L_\lambda^*$. For the  parameters considered here, Eq.~\eqref{eq:critical} predicts the critical flagella axial length $L_{\lambda,crit}^*\approx 3.25$ above which the bound state is unstable. This value obtained from Eq.~\eqref{eq:critical} is shown as purple colored vertical line in Fig.~\ref{fig:results}(c), showing excellent agreement with that obtained from numerically integrating the coupled Eqs.~\eqref{eq:theoryforce} and \eqref{eq:theorytorque}.

\subsection{Scaling arguments}

The computational and theoretical results showing a  fluid-structure interaction instability leading to the dynamic bound state of bacteria can be rationalised using simple scaling arguments.  First, let us consider the force balance on the cell at steady state. The typical magnitude of the propulsive force acting on the cell in the direction parallel to the surface is  $\sim f\cos \theta L_\lambda$ while the typical drag on the cell is $\sim \mu U (a+L_\lambda) $, with logarithmic corrections arising from lubrication cell-wall interactions~\cite{kimbook} that we ignore from a scaling point of view. Balancing these forces leads to the first scaling result, $\mu U (a+L_\lambda) \sim f\cos \theta L_\lambda$. 

The second scaling identity arises from the balance of torques. The cell body translates along the surface and as a result experiences a torque (which makes it roll and aligns the flagellum perpendicular to the surface) of magnitude $\sim \mu U  a^2$, again with logarithmic corrections that we ignore. In addition, the propelling flagellum experiences a viscous torque that acts to sweep it behind the cell body due to the swimming motion of the whole cell. This torque scales as 
$\sim \mu U L_\lambda (a+L_\lambda)\sin \theta$ when measured from the center of the cell body. The $\sin \theta$ term appears because only the flow perpendicular to the flagellum induces a viscous torque. Balancing these two torques leads to the second identity, $\mu U  a^2 \sim \mu U L_\lambda (a+L_\lambda)\sin \theta$. 

Combining the two relationships obtained from the force and torque balances, we obtain the identity 
\begin{equation}
a^2 \cos\theta\sim  L_\lambda(a+L_\lambda) \sin \theta \cos\theta.
\end{equation}
 This equation has two solutions for the angle $\theta$. The first being $\cos\theta = 0$, corresponding to the surface-bound state with the flagellum pointing into the surface, and thereby, producing no motion. The second solution is 
given by $\sin\theta =  a^2/(aL_\lambda + L_\lambda^2)$, corresponding to a state where the flagellum is tilted, so the cell swims parallel to the wall and undergoes circular swimming. It is evident that this equation has no solution if $L_\lambda$ is too small, explaining the stability of the surface-bound state for cells with small flagella, and the existence of a new state at a critical value of the axial flagellum length, $L_\lambda$, as seen in Fig.~\ref{fig:results}. Furthermore, the scaling arguments predict  $\theta \to 0$ for large values of $L_\lambda$, in agreement with our numerical simulations and theoretical model.
 
\section{Conclusion}\label{sec:conclusion}

In this article, we have proposed a model explaining the ability of spherical shaped motile \textit{T.~majus} bacterial cells to become surface bound. We have shown how a large ratio of cell body size to flagellar length can cause transition from circular swimming along the surface to a bound state. These results have  significant implications on the initial stages of formation of the white veil, an approximately \SI{0.5}{\milli\meter}-thick elastic porous medium that is the natural habitat of \textit{T.~majus} cells \cite{fenchel1998}. Our results are in agreement with the experiments in Ref.~\cite{petroff2018}, where about 90\% of cells were observed to be in the surface-bound state while the rest were observed to swim in circular paths. We attribute the small fraction of swimming cells to natural variations in flagellar lengths amongst cell populations. 

Our results also suggest that bacteria can pump fluid normal to a rigid surface possibly increasing nutrient advection, as opposed to previous studies where helical pumping due to stuck bacteria occurred parallel to the surface \cite{gao2015,dauparas2018}. 
However, many other questions related to \textit{T.~majus} locomotion dynamics  remain unanswered. How are these cells able to swim so fast? Do they possess an elastic hook like the well-studied 
\textit{E.~coli} and if yes, how do hooks affect the propulsion of these microorganisms \cite{riley2018,ishimoto2019}? A recent hydrodynamic study has proposed that fast swimming of \textit{T.~majus} cells occurs due to multiple rotating flagella present on its surface \cite{belovs2016}. However, it is not clear as to why, or how, they must all point in the same direction. Detailed experimental work is needed in the future to measure the rotation speed of the flagella, similar to that done for \textit{E.~coli} \cite{darnton2007}, as it is possible that fast cell swimming is a result of fast flagellar rotation. It is instructive to note that, while we have modeled the cell as having a single flagellum, the presence of multiple propelling flagella   can create wobbling effects, as  seen for \textit{E. coli}~\cite{darnton2007,hyon2012,bianchi2017}, which may impact   the stability of the surface-bound state.

Finally, we note that the surface-binding phenomena have been observed with at least two different bacterial species, namely, \textit{E. coli}~\cite{drescher2011,dominick2018} and \textit{Serratia marcescens}~\cite{chen2015}, both 10 times smaller than \textit{T.~majus} cells but possessing an elongated sphero-cylindrical cell body. Furthermore, in the case of \textit{Serratia marcescens}, the cells became bound to an air-liquid interface in the same way as \textit{T.~majus} cells, suggesting that the surface binding mechanism described here is not just restricted to solid surfaces.

Note. Recently, the authors were made aware of work by K. Ishimoto similar to that presented here~\cite{ishimoto2019b}.


\section*{Acknowledgements}
We thank A.~P.~Petroff for helpful discussions. This project has received funding from the European Research Council under the European Union's Horizon 2020 research and innovation programme (Grant No. 682754 to E.~L.). 

\appendix
\onecolumngrid
\section{Slender-body theory operators and Stokeslet near a wall}\label{sec:sbtappendix}
In Eq.~(2) in the main text, the  local, $\bm{\Lambda}$, and non-local, $\bm{K}$, operators are given by~\cite{johnson1980}
\begin{align}
\bm{\Lambda}[\bm{f}_h](\bm{x}_0) &= [-c(\mathsfbi{I} + \bm{\hat{s}}\bm{\hat{s}}) + 2(\mathsfbi{I} - \bm{\hat{s}}\bm{\hat{s}})] \bcdot \bm{f}_h(\bm{x}_0), \label{operators1}  \\
\bm{K}[\bm{f}_h](\bm{x}_0) &= \int\displaylimits_{0}^{L} \left(\bm{f}_h(\bm{x})\bcdot \mathsfbi{G}(\bm{x},\bm{x}_0) - \frac{\mathsfbi{I} + \bm{\hat{s}}\bm{\hat{s}}}{|s-s^\prime|}\bcdot \bm{f}_h(\bm{x}_0)\right) \,\mathrm{d}s^\prime(\bm{x}),\label{operators2} 
\end{align}
where $c=\log{(\varepsilon^2 e)}$. The tensor $\mathsfbi{G}(\bm{x},\bm{x}_0)$ is the free space Green's function for Stokes equation also called the Oseen--Burgers tensor, representing fluid flow produced by a point force,
\begin{equation}\label{eq:oseenmathsfbi}
\mathsfbi{G}(\bm{x},\bm{x}_0) = \frac{\mathsfbi{I}}{|\bm{x}-\bm{x}_0|} + \frac{(\bm{x}-\bm{x}_0)(\bm{x}-\bm{x}_0)}{|\bm{x}-\bm{x}_0|^3},
\end{equation}
where $\mathsfbi{I}$ is the $3\times3$ identity tensor. To account for the no slip velocity condition on the wall, let us consider a Stokeslet placed at a distance $h$ above the wall at $z=0$ such that its location is $(y_1,y_2,h)$. The image singularities are then accordingly located below the wall at $(y_1,y_2,-h)$. The Green's function~\cite{blake1974} due to the stokeslet at an evaluation point $(x_1,x_2,x_3)$ is
\begin{equation}
G_{ij}^w = \frac{\delta_{ij} + \hat{r}_i\hat{r}_j}{r} -\frac{\delta_{ij} + \hat{R}_i\hat{R}_j}{R} + 2h \Delta_{jk}\frac{\partial}{\partial R_k}\left(\frac{h\hat{R}_i}{R^2} - \frac{\delta_{i3} + \hat{R}_i\hat{R}_3 }{R}\right),
\end{equation}
where the vector pointing from the stokeslet location to the evaluation point is $r_i=(x_1-y_1,x_2-y_2,x_3-h)$, the vector pointing from the image location to the evaluation point is $R_i=(x_1-y_1,x_2-y_2,x_3+h)$ and the matrix $\Delta_{jk}$ is
\begin{equation}\label{eq:matrixdeltajk}
\Delta_{jk} =
\begin{bmatrix}
1 & 0 & 0 \\
0 & 1 & 0 \\
0 & 0 & -1 \\
\end{bmatrix}.
\end{equation}
Note that in this slender body theory formulation, the cross-sectional radius of the  body varies slowly as $r(s) = 2\varepsilon \sqrt{s(L-s)}$ where $\varepsilon = r(L/2)/L$, ensuring algebraically-accurate results. The cross-sectional radius at the midpoint $s=L/2$ is taken to be equal to the radius of the flagellum, i.e.~$r(L/2) =\rho$ . The kernel in Eq.~\eqref{operators2} becomes formally singular when $s=s^\prime$ and this singularity is removed by regularising the integral~\cite{tornberg2004}. 

\section{Suitability of boundary element method at resolving small gaps between surfaces}\label{sec:bemtest}

As mentioned in the main text, the boundary element method (BEM) becomes increasingly untenable as the distance between the sphere's surface and the plane wall decreases. The discretization error due to BEM using a single layer potential formulation~\cite{pozrikidis2002} can be computed by comparing the exact solution for the resistance matrix of the sphere near a wall in bispherical coordinates~\cite{jeffery1915,stimson1926,brenner1961,maude1961,dean1963,oneill1964}. For example, a spherical uniform mesh of 5120 elements, generates a grid of size $\Delta x \approx 0.075$, which prescribes the minimum distance between the sphere surface and the wall to be $\delta \approx \Delta x/2 = 0.0375$ while maintaining reasonable accuracy. However, these distances are two orders of magnitude higher than the desired gap height of $\delta\approx 0.0001$. This issue can be circumvented slightly by generating a non-uniform mesh where a higher number of nodes are generated on the sphere's surface closer to the wall~\cite{loewenberg1996}. 

\begin{figure*}
    \centering
    \includegraphics[width=0.95\linewidth]{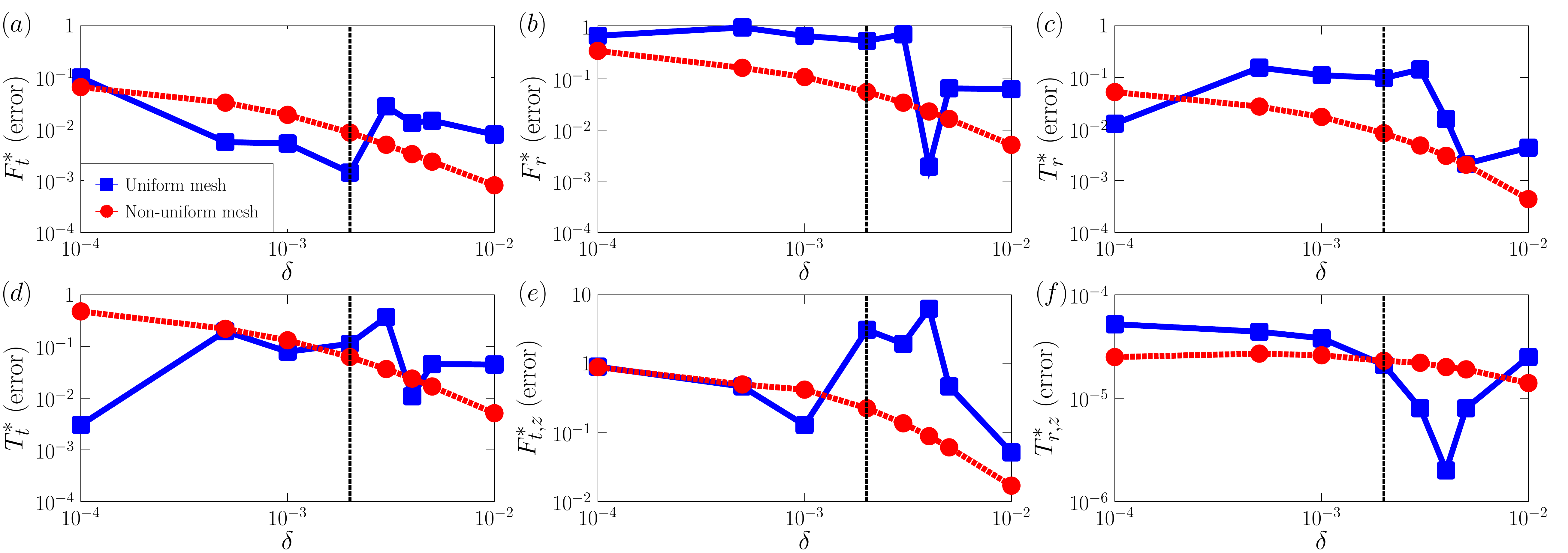}
    \caption{Relative errors of the elements of the resistance matrix of a sphere near a wall computed by boundary element method for a uniform and non-uniform mesh when compared with exact solutions using bispherical coordinates.}
    \label{fig:meshtest}
\end{figure*}

Denoting the values at the center with the subscript `C' and the values at the junction without any subscript, the resistance matrix of a sphere near a wall about its center $\bm{x}_C$ is written as
\begin{equation}\label{eq:RMcenter}
\begin{bmatrix} 
\bm{F}_C \\ 
\bm{T}_C
\end{bmatrix}
 =
  \begin{bmatrix}
   \mathsfbi{A}_C & \mathsfbi{B}_C \\
   \mathsfbi{B}^T_C & \mathsfbi{C}_C \\
  \end{bmatrix}
  \begin{bmatrix} 
\bm{U}_C \\ 
\bm{\Omega}_C 
\end{bmatrix}.
\end{equation}
Assuming the wall to be parallel to the $x-y$ plane, the resistance matrix can be expanded as,
\begin{equation}\label{eq:RMexpanded}
\begin{bmatrix} 
F_x \\
F_y \\
F_z \\
T_x \\
T_y \\
T_z \\
\end{bmatrix}_C
 =
  -\begin{bmatrix}
   6\pi\mu a F_{t}^* & 0 & 0 & 0 & 6\pi\mu a^2 F_{r}^* & 0 \\
   0 & 6\pi\mu a F_{t}^* & 0 & -6\pi\mu a^2 F_{r}^* & 0 & 0 \\
   0 & 0 & 6\pi\mu a F_{t,z}^* & 0 & 0 & 0 \\
   0 & -8\pi\mu a^2 T_{t}^* & 0 & 8\pi\mu a^3 T_{r}^* & 0 & 0 \\
   8\pi\mu a^2 T_{t}^* & 0 & 0 & 0 & 8\pi\mu a^3 T_{r}^* & 0 \\
   0 & 0 & 0 & 0 & 0 & 8\pi\mu a^3 T_{r,z}^* \\
  \end{bmatrix}
\begin{bmatrix} 
U_x \\
U_y \\
U_z \\
\Omega_x \\
\Omega_y \\
\Omega_z \\
\end{bmatrix}_C.
\end{equation}
The relative errors, $1 - F_{\mathrm{bem}}/F_{\mathrm{exact}}$, of various elements of the resistance matrix $F_{t}^*$, $F_{r}^*$, $T_{r}^*$, $T_{t}^*$, $F_{t,z}^*$ and $T_{r,z}^*$ for a uniform mesh and a non-uniform mesh obtained by BEM are shown in Fig.~\ref{fig:meshtest}. The minimum distance between the sphere surface and the wall that can be resolved with the non-uniform mesh is $\delta \approx \Delta x_{\mathrm{min}}/2 = 0.002$, shown in black dashed line in Fig.~\ref{fig:meshtest}. The uniform mesh performs quite poorly compared to the non-uniform mesh, particularly so for the coefficient $F_{t,z}^*$. In all the cases, the relative error with the uniform mesh oscillates rather than varies smoothly with the gap height $\delta$ when compared with the non-uniform mesh. The relative errors are even higher for a higher grid size or lower number of surface nodes, not shown here for brevity. The numerical error can be reduced by using a high order quadrature for integrating the kernel $\mathsfbi{G}$ in the boundary integral equation, however, the number of grid points  required for performing dynamical simulations still remains very high. As a consequence, we cannot use the boundary element method to accurately resolve hydrodynamics of a body extremely close to a surface, and instead use the method outlined in Appendix~\ref{sec:rmmatrices}.

\section{Far-field and lubrication limits compared with exact solutions}\label{sec:rmmatrices}
In order to circumvent the issue described in Appendix \ref{sec:bemtest}, we use analytical results in the far-field and lubrication limits. The coefficients appearing in the resistance matrix \eqref{eq:RMexpanded} can be computed exactly in bispherical coordinates. However, it requires solving a linear system whose size increases with decreasing gap height $\delta$. Alternatively, they can be derived using lubrication approximation and far-field approximation. 

\begin{figure*}
    \centering
    \includegraphics[width=0.95\linewidth]{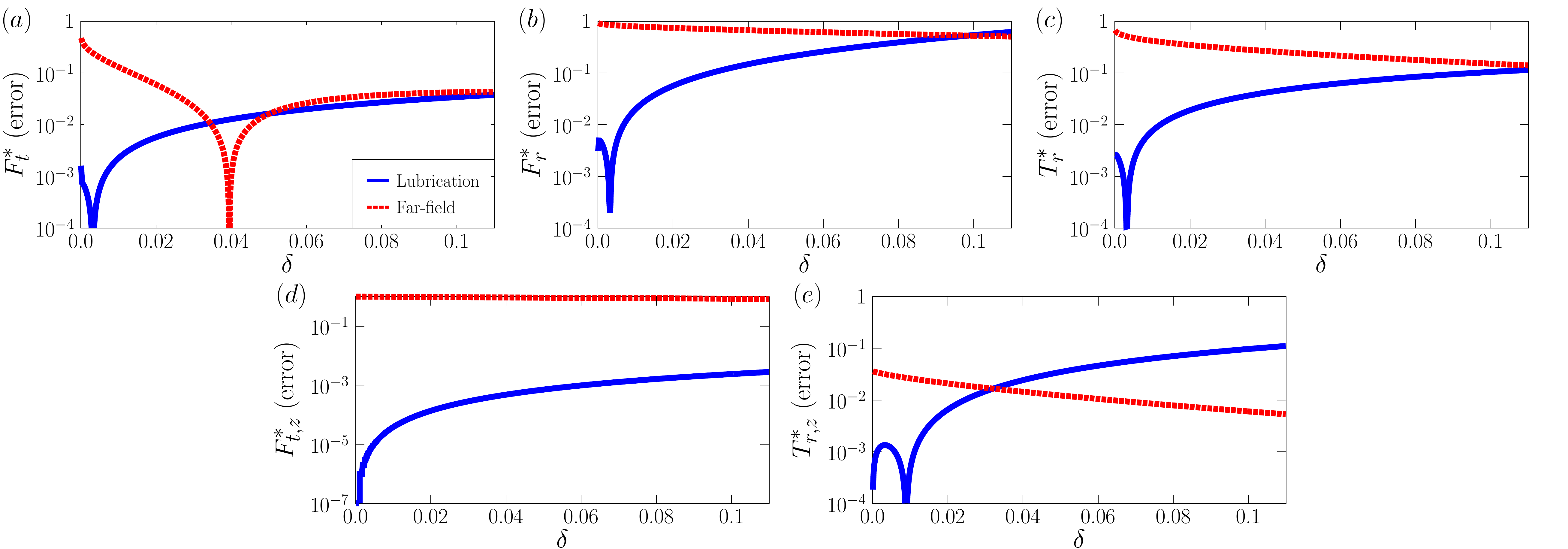}
    \caption{Relative errors of the elements of the resistance matrix of a sphere near a wall obtained using lubrication and far-field theory when compared with exact solutions using bispherical coordinates.}
    \label{fig:exactnearfar}
\end{figure*}
 
For a given distance from the wall, we have compared these coefficients with the exact solutions and determined a cut-off distance accordingly. The relative errors in the coefficients obtained by far-field and lubrication theory when compared to exact solutions are shown in Fig.~\ref{fig:exactnearfar}. Specifically, we use lubrication formulas~\cite{brenner1961,goldman1967} listed below for $F_{t}^*$, $F_{r}^*$, $T_{t}^*$, $T_{r}^*$ and $F_{t,z}^*$ when $\delta<0.1$, and $T_{r,z}^*$ when $\delta<0.03$,
\begin{align}
\begin{split}
&F_{t}^* =-\frac{8}{15}\ln (\delta) + 0.9588,~ F_{r}^* = \frac{2}{15}\ln (\delta) + 0.2526, ~T_{r}^* = -\frac{2}{5}\ln (\delta) + 0.3817, \\
&T_{t}^* = \frac{1}{10}\ln (\delta) + 0.1895, ~F_{t,z}^* = \frac{1}{\delta}\left[1 + \frac{1}{5}\delta\ln\left(\frac{1}{\delta}\right) + k\delta\right], ~T_{r,z}^* = \zeta(3) + 3\left(\frac{\pi^2}{6}-1\right)\delta,
\end{split}
\end{align}
where $k=0.971624$ and $\zeta$ is Riemann's function with $\zeta(3) \simeq 1.20206$. For the complementary values of $\delta$, we use far-field analytical solutions,
\begin{align}
\begin{split}
&F_{t}^* = \left[1 - \frac{9}{16d} + \frac{1}{8d^3} - \frac{45}{256d^4} - \frac{1}{16d^5}\right]^{-1},~F_{r}^* = -\frac{1}{8d^4}\left(1-\frac{3}{8d}\right), ~T_{r}^* = 1+\frac{5}{16d^3}, \\
&T_{t}^* =  -\frac{3}{32d^4}\left(1-\frac{3}{8d}\right), ~F_{t,z}^* = \left[1 - \frac{9}{16d} + \frac{1}{8d^3}\right]^{-1}, ~T_{r,z}^* = \left[1 - \frac{1}{8d^3} - \frac{3}{256d^8}\right]^{-1}.
\end{split}
\end{align}
Note that as $F_{r}^*/T_{t}^* = 3/4$, the relative error plot for $T_{t}^*$ is identical to  $F_{r}^*$ and has not been shown in Fig.~\ref{fig:exactnearfar}. If the translation and rotation are only along the $x$ and $y$ direction, respectively (as in the theoretical rod model) Eq.~\eqref{eq:RMexpanded} reduces to Eq.~(11) in the main text. 

Lastly, for the numerical model in \S IV of the main text, we need a relationship between forces, torques and velocities at the junction, provided by rigid body dynamics,
\begin{equation}\label{eq:rigidbody}
\begin{split}
& \bm{F}_C = \bm{F},~\bm{T}_C = \bm{T} + \bm{r} \btimes \bm{F},~\bm{F}_C = \mathsfbi{A}_C \bcdot \bm{U}_C, \\
& \bm{\Omega}_C = \bm{\Omega},~\bm{U}_C = \bm{U} + \bm{r} \btimes \bm{\Omega},~\bm{T}_C = \mathsfbi{C}_C \bcdot \bm{\Omega}_C,
\end{split}
\end{equation}
where $\bm{r} = \bm{x}_J - \bm{x}_C$. Manipulating the equations in \eqref{eq:rigidbody}, we get the desired resistance matrix computed about the junction $\bm{x}_J$,
\begin{equation}\label{eq:RMjunction}
\begin{bmatrix} 
\bm{F} \\ 
\bm{T}
\end{bmatrix}
 =
  \begin{bmatrix}
   \mathsfbi{A} & \mathsfbi{B} \\
   \mathsfbi{B}^T & \mathsfbi{C} \\
  \end{bmatrix}
  \begin{bmatrix} 
\bm{U} \\ 
\bm{\Omega}
\end{bmatrix},
\end{equation}
where $\mathsfbi{A} =  \mathsfbi{A}_C$, $\mathsfbi{B} =  \mathsfbi{B}_C + \mathsfbi{A}_C \bcdot \mathsfbi{R}$, and $\mathsfbi{C} = \mathsfbi{C}_C + \mathsfbi{B}^T_C \bcdot \mathsfbi{R} + \mathsfbi{R}^T \bcdot \mathsfbi{B}_C + \mathsfbi{R} \bcdot \mathsfbi{A}_C \bcdot \mathsfbi{R}^T$. We have introduced the cross-product matrix $\mathsfbi{R}$, so that $\bm{r} \btimes \bm{v} = \mathsfbi{R}\bcdot \bm{v}$, where $\bm{v}$ is any vector,
\begin{align}
R_{ik} = \epsilon_{ijk}r_j=
\begin{bmatrix} 
0 & -r_3 & r_2 \\
r_3 & 0 & -r_1 \\
-r_2 & r_1 & 0
\end{bmatrix}.
\end{align}

\section{Resistance coefficients of a rod}\label{sec:resistancerod}
The drag coefficients used in the the rod model in \S IV of the main text are obtained by assuming them to be slender prolate spheroids~\cite{chwang1975},
\begin{align}
c_t = \frac{2\pi\mu}{\log(L/\rho) - 1/2}, ~c_n = \frac{4\pi\mu}{\log(L/\rho) + 1/2}.
\end{align}

\begin{figure*}
    \centering
    \includegraphics[width=0.95\linewidth]{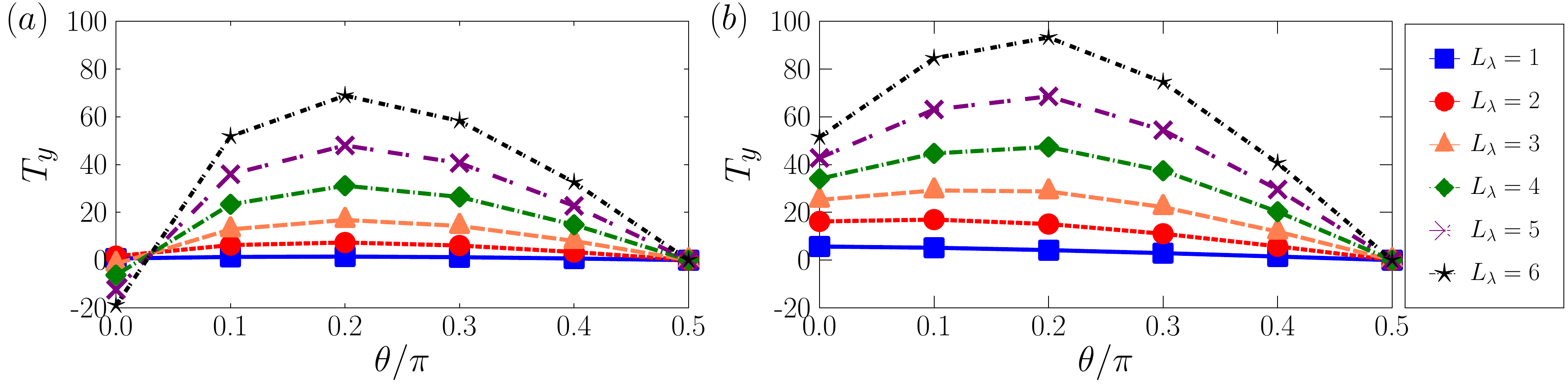}
    \caption{Hydrodynamic torque $T_y$ acting on a rotating helix measured about its tapered end for two different heights (a) $d=1.005$, and (b) $d=0.25$ above the wall plotted against various tilt angles $\theta$ for various flagellar axial lengths $L_\lambda$. See the schematic diagram in Fig.~2 for an illustration of geometrical parameters and coordinate axis definition.}
    \label{fig:rotatinghelix}
\end{figure*}

\section{Torque on a rotating helix near a wall}\label{sec:torquehelix}
A rotating helix placed in a semi-infinite fluid medium experiences a hydrodynamic torque that is mostly attractive except when the helix is parallel or nearly parallel and placed above a certain height above the wall (Fig.~\ref{fig:rotatinghelix}). As mentioned in the main text, this attractive torque tends to align the helix parallel to the wall.

\newpage

\bibliography{papers} 

\begin{thebibliography}{48}%
\makeatletter
\providecommand \@ifxundefined [1]{%
 \@ifx{#1\undefined}
}%
\providecommand \@ifnum [1]{%
 \ifnum #1\expandafter \@firstoftwo
 \else \expandafter \@secondoftwo
 \fi
}%
\providecommand \@ifx [1]{%
 \ifx #1\expandafter \@firstoftwo
 \else \expandafter \@secondoftwo
 \fi
}%
\providecommand \natexlab [1]{#1}%
\providecommand \enquote  [1]{``#1''}%
\providecommand \bibnamefont  [1]{#1}%
\providecommand \bibfnamefont [1]{#1}%
\providecommand \citenamefont [1]{#1}%
\providecommand \href@noop [0]{\@secondoftwo}%
\providecommand \href [0]{\begingroup \@sanitize@url \@href}%
\providecommand \@href[1]{\@@startlink{#1}\@@href}%
\providecommand \@@href[1]{\endgroup#1\@@endlink}%
\providecommand \@sanitize@url [0]{\catcode `\\12\catcode `\$12\catcode
  `\&12\catcode `\#12\catcode `\^12\catcode `\_12\catcode `\%12\relax}%
\providecommand \@@startlink[1]{}%
\providecommand \@@endlink[0]{}%
\providecommand \url  [0]{\begingroup\@sanitize@url \@url }%
\providecommand \@url [1]{\endgroup\@href {#1}{\urlprefix }}%
\providecommand \urlprefix  [0]{URL }%
\providecommand \Eprint [0]{\href }%
\providecommand \doibase [0]{http://dx.doi.org/}%
\providecommand \selectlanguage [0]{\@gobble}%
\providecommand \bibinfo  [0]{\@secondoftwo}%
\providecommand \bibfield  [0]{\@secondoftwo}%
\providecommand \translation [1]{[#1]}%
\providecommand \BibitemOpen [0]{}%
\providecommand \bibitemStop [0]{}%
\providecommand \bibitemNoStop [0]{.\EOS\space}%
\providecommand \EOS [0]{\spacefactor3000\relax}%
\providecommand \BibitemShut  [1]{\csname bibitem#1\endcsname}%
\let\auto@bib@innerbib\@empty
\bibitem [{\citenamefont {Schulz}\ and\ \citenamefont
  {J{\o}rgensen}(2001)}]{schulz2001}%
  \BibitemOpen
  \bibfield  {author} {\bibinfo {author} {\bibfnamefont {H.~N.}\ \bibnamefont
  {Schulz}}\ and\ \bibinfo {author} {\bibfnamefont {B.~B.}\ \bibnamefont
  {J{\o}rgensen}},\ }\bibfield  {title} {\enquote {\bibinfo {title} {Big
  bacteria},}\ }\href@noop {} {\bibfield  {journal} {\bibinfo  {journal} {Annu.
  Rev. Microbiol.}\ }\textbf {\bibinfo {volume} {55}},\ \bibinfo {pages}
  {105--137} (\bibinfo {year} {2001})}\BibitemShut {NoStop}%
\bibitem [{\citenamefont {Schulz}\ \emph {et~al.}(1999)\citenamefont {Schulz},
  \citenamefont {Brinkhoff}, \citenamefont {Ferdelman}, \citenamefont
  {Marin{\'e}}, \citenamefont {Teske},\ and\ \citenamefont
  {J{\o}rgensen}}]{schulz1999}%
  \BibitemOpen
  \bibfield  {author} {\bibinfo {author} {\bibfnamefont {H.~N.}\ \bibnamefont
  {Schulz}}, \bibinfo {author} {\bibfnamefont {T.}~\bibnamefont {Brinkhoff}},
  \bibinfo {author} {\bibfnamefont {T.~G.}\ \bibnamefont {Ferdelman}}, \bibinfo
  {author} {\bibfnamefont {M.~H.}\ \bibnamefont {Marin{\'e}}}, \bibinfo
  {author} {\bibfnamefont {A.}~\bibnamefont {Teske}}, \ and\ \bibinfo {author}
  {\bibfnamefont {B.~B.}\ \bibnamefont {J{\o}rgensen}},\ }\bibfield  {title}
  {\enquote {\bibinfo {title} {Dense populations of a giant sulfur bacterium in
  {N}amibian shelf sediments},}\ }\href@noop {} {\bibfield  {journal} {\bibinfo
   {journal} {Science}\ }\textbf {\bibinfo {volume} {284}},\ \bibinfo {pages}
  {493--495} (\bibinfo {year} {1999})}\BibitemShut {NoStop}%
\bibitem [{\citenamefont {Berg}(2008)}]{berg2008}%
  \BibitemOpen
  \bibfield  {author} {\bibinfo {author} {\bibfnamefont {H.~C.}\ \bibnamefont
  {Berg}},\ }\href@noop {} {\emph {\bibinfo {title} {\textit{E. coli} in
  Motion}}}\ (\bibinfo  {publisher} {Springer Science \& Business Media},\
  \bibinfo {year} {2008})\BibitemShut {NoStop}%
\bibitem [{\citenamefont {Hinze}(1913)}]{hinze1913}%
  \BibitemOpen
  \bibfield  {author} {\bibinfo {author} {\bibfnamefont {G.}~\bibnamefont
  {Hinze}},\ }\bibfield  {title} {\enquote {\bibinfo {title} {Beitr{\"a}ge zur
  {K}enntnis der farblosen {S}chwefelbakterien},}\ }\href@noop {} {\bibfield
  {journal} {\bibinfo  {journal} {Ber. Dtsch. Bot. Ges.}\ }\textbf {\bibinfo
  {volume} {31}},\ \bibinfo {pages} {189--202} (\bibinfo {year}
  {1913})}\BibitemShut {NoStop}%
\bibitem [{\citenamefont {Wirsen}\ and\ \citenamefont
  {Jannasch}(1978)}]{wirsen1978}%
  \BibitemOpen
  \bibfield  {author} {\bibinfo {author} {\bibfnamefont {C.~O.}\ \bibnamefont
  {Wirsen}}\ and\ \bibinfo {author} {\bibfnamefont {H.~W.}\ \bibnamefont
  {Jannasch}},\ }\bibfield  {title} {\enquote {\bibinfo {title} {Physiological
  and morphological observations on \textit{Thiovulum sp}.}}\ }\href@noop {}
  {\bibfield  {journal} {\bibinfo  {journal} {J. Bacteriol.}\ }\textbf
  {\bibinfo {volume} {136}},\ \bibinfo {pages} {765--774} (\bibinfo {year}
  {1978})}\BibitemShut {NoStop}%
\bibitem [{\citenamefont {Garcia-Pichel}(1989)}]{garcia1989}%
  \BibitemOpen
  \bibfield  {author} {\bibinfo {author} {\bibfnamefont {F.}~\bibnamefont
  {Garcia-Pichel}},\ }\bibfield  {title} {\enquote {\bibinfo {title} {Rapid
  bacterial swimming measured in swarming cells of \textit{Thiovulum majus}},}\
  }\href@noop {} {\bibfield  {journal} {\bibinfo  {journal} {J. Bacteriol.}\
  }\textbf {\bibinfo {volume} {171}},\ \bibinfo {pages} {3560--3563} (\bibinfo
  {year} {1989})}\BibitemShut {NoStop}%
\bibitem [{\citenamefont {Fenchel}\ and\ \citenamefont
  {Thar}(2004)}]{fenchel2004}%
  \BibitemOpen
  \bibfield  {author} {\bibinfo {author} {\bibfnamefont {Tom}\ \bibnamefont
  {Fenchel}}\ and\ \bibinfo {author} {\bibfnamefont {Roland}\ \bibnamefont
  {Thar}},\ }\bibfield  {title} {\enquote {\bibinfo {title} {``{C}andidatus
  {O}vobacter propellens'': a large conspicuous prokaryote with an unusual
  motility behaviour},}\ }\href@noop {} {\bibfield  {journal} {\bibinfo
  {journal} {FEMS Microbiol. Ecol.}\ }\textbf {\bibinfo {volume} {48}},\
  \bibinfo {pages} {231--238} (\bibinfo {year} {2004})}\BibitemShut {NoStop}%
\bibitem [{\citenamefont {Fenchel}\ and\ \citenamefont
  {Glud}(1998)}]{fenchel1998}%
  \BibitemOpen
  \bibfield  {author} {\bibinfo {author} {\bibfnamefont {T.}~\bibnamefont
  {Fenchel}}\ and\ \bibinfo {author} {\bibfnamefont {R.~N.}\ \bibnamefont
  {Glud}},\ }\bibfield  {title} {\enquote {\bibinfo {title} {Veil architecture
  in a sulphide-oxidizing bacterium enhances countercurrent flux},}\
  }\href@noop {} {\bibfield  {journal} {\bibinfo  {journal} {Nature}\ }\textbf
  {\bibinfo {volume} {394}},\ \bibinfo {pages} {367} (\bibinfo {year}
  {1998})}\BibitemShut {NoStop}%
\bibitem [{\citenamefont {Petroff}\ \emph
  {et~al.}(2015{\natexlab{a}})\citenamefont {Petroff}, \citenamefont {Pasulka},
  \citenamefont {Soplop}, \citenamefont {Wu},\ and\ \citenamefont
  {Libchaber}}]{petroff2015a}%
  \BibitemOpen
  \bibfield  {author} {\bibinfo {author} {\bibfnamefont {A.~P.}\ \bibnamefont
  {Petroff}}, \bibinfo {author} {\bibfnamefont {A.~L.}\ \bibnamefont
  {Pasulka}}, \bibinfo {author} {\bibfnamefont {N.}~\bibnamefont {Soplop}},
  \bibinfo {author} {\bibfnamefont {X.-L.}\ \bibnamefont {Wu}}, \ and\ \bibinfo
  {author} {\bibfnamefont {A.}~\bibnamefont {Libchaber}},\ }\bibfield  {title}
  {\enquote {\bibinfo {title} {Biophysical basis for convergent evolution of
  two veil-forming microbes},}\ }\href@noop {} {\bibfield  {journal} {\bibinfo
  {journal} {R. Soc. Open Sci.}\ }\textbf {\bibinfo {volume} {2}},\ \bibinfo
  {pages} {150437} (\bibinfo {year} {2015}{\natexlab{a}})}\BibitemShut
  {NoStop}%
\bibitem [{\citenamefont {De~Boer}\ \emph {et~al.}(1961)\citenamefont
  {De~Boer}, \citenamefont {La~Riviere},\ and\ \citenamefont
  {Houwink}}]{deboer1961}%
  \BibitemOpen
  \bibfield  {author} {\bibinfo {author} {\bibfnamefont {W.~E.}\ \bibnamefont
  {De~Boer}}, \bibinfo {author} {\bibfnamefont {J.~W.~M.}\ \bibnamefont
  {La~Riviere}}, \ and\ \bibinfo {author} {\bibfnamefont {A.~L.}\ \bibnamefont
  {Houwink}},\ }\bibfield  {title} {\enquote {\bibinfo {title} {Observations on
  the morphology of \textit{Thiovulum majus} {H}inze},}\ }\href@noop {}
  {\bibfield  {journal} {\bibinfo  {journal} {Antonie van Leeuwenhoek J.
  Microbiol. Serol.}\ }\textbf {\bibinfo {volume} {27}},\ \bibinfo {pages}
  {447--456} (\bibinfo {year} {1961})}\BibitemShut {NoStop}%
\bibitem [{\citenamefont {Fenchel}(1994)}]{fenchel1994}%
  \BibitemOpen
  \bibfield  {author} {\bibinfo {author} {\bibfnamefont {T.}~\bibnamefont
  {Fenchel}},\ }\bibfield  {title} {\enquote {\bibinfo {title} {Motility and
  chemosensory behaviour of the sulphur bacterium \textit{Thiovulum majus}},}\
  }\href@noop {} {\bibfield  {journal} {\bibinfo  {journal} {Microbiology}\
  }\textbf {\bibinfo {volume} {140}},\ \bibinfo {pages} {3109--3116} (\bibinfo
  {year} {1994})}\BibitemShut {NoStop}%
\bibitem [{\citenamefont {Petroff}\ \emph
  {et~al.}(2015{\natexlab{b}})\citenamefont {Petroff}, \citenamefont {Wu},\
  and\ \citenamefont {Libchaber}}]{petroff2015b}%
  \BibitemOpen
  \bibfield  {author} {\bibinfo {author} {\bibfnamefont {A.~P.}\ \bibnamefont
  {Petroff}}, \bibinfo {author} {\bibfnamefont {X.-L.}\ \bibnamefont {Wu}}, \
  and\ \bibinfo {author} {\bibfnamefont {A.}~\bibnamefont {Libchaber}},\
  }\bibfield  {title} {\enquote {\bibinfo {title} {Fast-moving bacteria
  self-organize into active two-dimensional crystals of rotating cells},}\
  }\href@noop {} {\bibfield  {journal} {\bibinfo  {journal} {Phys. Rev. Lett.}\
  }\textbf {\bibinfo {volume} {114}},\ \bibinfo {pages} {158102} (\bibinfo
  {year} {2015}{\natexlab{b}})}\BibitemShut {NoStop}%
\bibitem [{\citenamefont {Petroff}\ and\ \citenamefont
  {Libchaber}(2018)}]{petroff2018}%
  \BibitemOpen
  \bibfield  {author} {\bibinfo {author} {\bibfnamefont {A.~P.}\ \bibnamefont
  {Petroff}}\ and\ \bibinfo {author} {\bibfnamefont {A.}~\bibnamefont
  {Libchaber}},\ }\bibfield  {title} {\enquote {\bibinfo {title} {Nucleation of
  rotating crystals by \textit{Thiovulum majus} bacteria},}\ }\href@noop {}
  {\bibfield  {journal} {\bibinfo  {journal} {New J. Phys.}\ }\textbf {\bibinfo
  {volume} {20}},\ \bibinfo {pages} {015007} (\bibinfo {year}
  {2018})}\BibitemShut {NoStop}%
\bibitem [{\citenamefont {Berg}\ and\ \citenamefont {Turner}(1990)}]{berg1990}%
  \BibitemOpen
  \bibfield  {author} {\bibinfo {author} {\bibfnamefont {H.~C.}\ \bibnamefont
  {Berg}}\ and\ \bibinfo {author} {\bibfnamefont {L.}~\bibnamefont {Turner}},\
  }\bibfield  {title} {\enquote {\bibinfo {title} {Chemotaxis of bacteria in
  glass capillary arrays. \textit{Escherichia coli}, motility, microchannel
  plate, and light scattering},}\ }\href@noop {} {\bibfield  {journal}
  {\bibinfo  {journal} {Biophys. J.}\ }\textbf {\bibinfo {volume} {58}},\
  \bibinfo {pages} {919--930} (\bibinfo {year} {1990})}\BibitemShut {NoStop}%
\bibitem [{\citenamefont {Ramia}\ \emph {et~al.}(1993)\citenamefont {Ramia},
  \citenamefont {Tullock},\ and\ \citenamefont {Phan-Thien}}]{ramia1993}%
  \BibitemOpen
  \bibfield  {author} {\bibinfo {author} {\bibfnamefont {M.}~\bibnamefont
  {Ramia}}, \bibinfo {author} {\bibfnamefont {D.~L.}\ \bibnamefont {Tullock}},
  \ and\ \bibinfo {author} {\bibfnamefont {N.}~\bibnamefont {Phan-Thien}},\
  }\bibfield  {title} {\enquote {\bibinfo {title} {The role of hydrodynamic
  interaction in the locomotion of microorganisms},}\ }\href@noop {} {\bibfield
   {journal} {\bibinfo  {journal} {Biophys. J.}\ }\textbf {\bibinfo {volume}
  {65}},\ \bibinfo {pages} {755--778} (\bibinfo {year} {1993})}\BibitemShut
  {NoStop}%
\bibitem [{\citenamefont {Lauga}\ \emph {et~al.}(2006)\citenamefont {Lauga},
  \citenamefont {DiLuzio}, \citenamefont {Whitesides},\ and\ \citenamefont
  {Stone}}]{lauga2006}%
  \BibitemOpen
  \bibfield  {author} {\bibinfo {author} {\bibfnamefont {E.}~\bibnamefont
  {Lauga}}, \bibinfo {author} {\bibfnamefont {W.~R.}\ \bibnamefont {DiLuzio}},
  \bibinfo {author} {\bibfnamefont {G.~M.}\ \bibnamefont {Whitesides}}, \ and\
  \bibinfo {author} {\bibfnamefont {H.~A.}\ \bibnamefont {Stone}},\ }\bibfield
  {title} {\enquote {\bibinfo {title} {Swimming in circles: motion of bacteria
  near solid boundaries},}\ }\href@noop {} {\bibfield  {journal} {\bibinfo
  {journal} {Biophys. J.}\ }\textbf {\bibinfo {volume} {90}},\ \bibinfo {pages}
  {400--412} (\bibinfo {year} {2006})}\BibitemShut {NoStop}%
\bibitem [{\citenamefont {Lauga}\ and\ \citenamefont
  {Powers}(2009)}]{lauga2009}%
  \BibitemOpen
  \bibfield  {author} {\bibinfo {author} {\bibfnamefont {E.}~\bibnamefont
  {Lauga}}\ and\ \bibinfo {author} {\bibfnamefont {.~R.}\ \bibnamefont
  {Powers}},\ }\bibfield  {title} {\enquote {\bibinfo {title} {The
  hydrodynamics of swimming microorganisms},}\ }\href@noop {} {\bibfield
  {journal} {\bibinfo  {journal} {Rep. Prog. Phys.}\ }\textbf {\bibinfo
  {volume} {72}},\ \bibinfo {pages} {096601} (\bibinfo {year}
  {2009})}\BibitemShut {NoStop}%
\bibitem [{\citenamefont {Lauga}(2016)}]{lauga2016}%
  \BibitemOpen
  \bibfield  {author} {\bibinfo {author} {\bibfnamefont {E.}~\bibnamefont
  {Lauga}},\ }\bibfield  {title} {\enquote {\bibinfo {title} {Bacterial
  hydrodynamics},}\ }\href@noop {} {\bibfield  {journal} {\bibinfo  {journal}
  {Annu. Rev. Fluid Mech.}\ }\textbf {\bibinfo {volume} {48}},\ \bibinfo
  {pages} {105--130} (\bibinfo {year} {2016})}\BibitemShut {NoStop}%
\bibitem [{\citenamefont {Higdon}(1979{\natexlab{a}})}]{higdon1979a}%
  \BibitemOpen
  \bibfield  {author} {\bibinfo {author} {\bibfnamefont {J.~J.~L.}\
  \bibnamefont {Higdon}},\ }\bibfield  {title} {\enquote {\bibinfo {title} {The
  hydrodynamics of flagellar propulsion: helical waves},}\ }\href@noop {}
  {\bibfield  {journal} {\bibinfo  {journal} {J. Fluid Mech.}\ }\textbf
  {\bibinfo {volume} {94}},\ \bibinfo {pages} {331--351} (\bibinfo {year}
  {1979}{\natexlab{a}})}\BibitemShut {NoStop}%
\bibitem [{\citenamefont {Higdon}(1979{\natexlab{b}})}]{higdon1979b}%
  \BibitemOpen
  \bibfield  {author} {\bibinfo {author} {\bibfnamefont {J.~J.~L.}\
  \bibnamefont {Higdon}},\ }\bibfield  {title} {\enquote {\bibinfo {title} {A
  hydrodynamic analysis of flagellar propulsion},}\ }\href@noop {} {\bibfield
  {journal} {\bibinfo  {journal} {J. Fluid Mech.}\ }\textbf {\bibinfo {volume}
  {90}},\ \bibinfo {pages} {685--711} (\bibinfo {year}
  {1979}{\natexlab{b}})}\BibitemShut {NoStop}%
\bibitem [{\citenamefont {Johnson}(1980)}]{johnson1980}%
  \BibitemOpen
  \bibfield  {author} {\bibinfo {author} {\bibfnamefont {R.~E.}\ \bibnamefont
  {Johnson}},\ }\bibfield  {title} {\enquote {\bibinfo {title} {An improved
  slender-body theory for {S}tokes flow},}\ }\href@noop {} {\bibfield
  {journal} {\bibinfo  {journal} {J. Fluid Mech.}\ }\textbf {\bibinfo {volume}
  {99}},\ \bibinfo {pages} {411--431} (\bibinfo {year} {1980})}\BibitemShut
  {NoStop}%
\bibitem [{\citenamefont {Blake}\ and\ \citenamefont
  {Chwang}(1974)}]{blake1974}%
  \BibitemOpen
  \bibfield  {author} {\bibinfo {author} {\bibfnamefont {J.~R.}\ \bibnamefont
  {Blake}}\ and\ \bibinfo {author} {\bibfnamefont {A.~T.}\ \bibnamefont
  {Chwang}},\ }\bibfield  {title} {\enquote {\bibinfo {title} {Fundamental
  singularities of viscous flow},}\ }\href@noop {} {\bibfield  {journal}
  {\bibinfo  {journal} {J. Eng. Math.}\ }\textbf {\bibinfo {volume} {8}},\
  \bibinfo {pages} {23--29} (\bibinfo {year} {1974})}\BibitemShut {NoStop}%
\bibitem [{\citenamefont {Das}\ and\ \citenamefont {Lauga}(2018)}]{das2018}%
  \BibitemOpen
  \bibfield  {author} {\bibinfo {author} {\bibfnamefont {D.}~\bibnamefont
  {Das}}\ and\ \bibinfo {author} {\bibfnamefont {E.}~\bibnamefont {Lauga}},\
  }\bibfield  {title} {\enquote {\bibinfo {title} {Computing the motor torque
  of \textit{Escherichia coli}},}\ }\href@noop {} {\bibfield  {journal}
  {\bibinfo  {journal} {Soft matter}\ }\textbf {\bibinfo {volume} {14}},\
  \bibinfo {pages} {5955--5967} (\bibinfo {year} {2018})}\BibitemShut {NoStop}%
\bibitem [{\citenamefont {Smith}\ \emph {et~al.}(2009)\citenamefont {Smith},
  \citenamefont {Gaffney}, \citenamefont {Blake},\ and\ \citenamefont
  {Kirkman-Brown}}]{smith2009}%
  \BibitemOpen
  \bibfield  {author} {\bibinfo {author} {\bibfnamefont {D.~J.}\ \bibnamefont
  {Smith}}, \bibinfo {author} {\bibfnamefont {E.~A.}\ \bibnamefont {Gaffney}},
  \bibinfo {author} {\bibfnamefont {J.~R.}\ \bibnamefont {Blake}}, \ and\
  \bibinfo {author} {\bibfnamefont {J.~C.}\ \bibnamefont {Kirkman-Brown}},\
  }\bibfield  {title} {\enquote {\bibinfo {title} {Human sperm accumulation
  near surfaces: a simulation study},}\ }\href@noop {} {\bibfield  {journal}
  {\bibinfo  {journal} {J. Fluid Mech.}\ }\textbf {\bibinfo {volume} {621}},\
  \bibinfo {pages} {289--320} (\bibinfo {year} {2009})}\BibitemShut {NoStop}%
\bibitem [{\citenamefont {Kim}\ and\ \citenamefont {Karrila}(1991)}]{kimbook}%
  \BibitemOpen
  \bibfield  {author} {\bibinfo {author} {\bibfnamefont {S.}~\bibnamefont
  {Kim}}\ and\ \bibinfo {author} {\bibfnamefont {J.~S.}\ \bibnamefont
  {Karrila}},\ }\href@noop {} {\emph {\bibinfo {title} {Microhydrodynamics:
  {P}rinciples and {S}elected {A}pplications.}}}\ (\bibinfo  {publisher}
  {Butterworth-Heinemann},\ \bibinfo {address} {Boston, MA},\ \bibinfo {year}
  {1991})\BibitemShut {NoStop}%
\bibitem [{\citenamefont {Gao}\ \emph {et~al.}(2015)\citenamefont {Gao},
  \citenamefont {Li}, \citenamefont {Chen},\ and\ \citenamefont
  {Zhang}}]{gao2015}%
  \BibitemOpen
  \bibfield  {author} {\bibinfo {author} {\bibfnamefont {Z.}~\bibnamefont
  {Gao}}, \bibinfo {author} {\bibfnamefont {H.}~\bibnamefont {Li}}, \bibinfo
  {author} {\bibfnamefont {X.}~\bibnamefont {Chen}}, \ and\ \bibinfo {author}
  {\bibfnamefont {H.~P.}\ \bibnamefont {Zhang}},\ }\bibfield  {title} {\enquote
  {\bibinfo {title} {Using confined bacteria as building blocks to generate
  fluid flow},}\ }\href@noop {} {\bibfield  {journal} {\bibinfo  {journal} {Lab
  on a Chip}\ }\textbf {\bibinfo {volume} {15}},\ \bibinfo {pages} {4555--4562}
  (\bibinfo {year} {2015})}\BibitemShut {NoStop}%
\bibitem [{\citenamefont {Dauparas}\ \emph {et~al.}(2018)\citenamefont
  {Dauparas}, \citenamefont {Das},\ and\ \citenamefont {Lauga}}]{dauparas2018}%
  \BibitemOpen
  \bibfield  {author} {\bibinfo {author} {\bibfnamefont {J.}~\bibnamefont
  {Dauparas}}, \bibinfo {author} {\bibfnamefont {D.}~\bibnamefont {Das}}, \
  and\ \bibinfo {author} {\bibfnamefont {E.}~\bibnamefont {Lauga}},\ }\bibfield
   {title} {\enquote {\bibinfo {title} {Helical micropumps near surfaces},}\
  }\href@noop {} {\bibfield  {journal} {\bibinfo  {journal} {Biomicrofluidics}\
  }\textbf {\bibinfo {volume} {12}},\ \bibinfo {pages} {014108} (\bibinfo
  {year} {2018})}\BibitemShut {NoStop}%
\bibitem [{\citenamefont {Riley}\ \emph {et~al.}(2018)\citenamefont {Riley},
  \citenamefont {Das},\ and\ \citenamefont {Lauga}}]{riley2018}%
  \BibitemOpen
  \bibfield  {author} {\bibinfo {author} {\bibfnamefont {E.~E.}\ \bibnamefont
  {Riley}}, \bibinfo {author} {\bibfnamefont {D.}~\bibnamefont {Das}}, \ and\
  \bibinfo {author} {\bibfnamefont {E.}~\bibnamefont {Lauga}},\ }\bibfield
  {title} {\enquote {\bibinfo {title} {Swimming of peritrichous bacteria is
  enabled by an elastohydrodynamic instability},}\ }\href@noop {} {\bibfield
  {journal} {\bibinfo  {journal} {Sci. Rep.}\ }\textbf {\bibinfo {volume} {8}}
  (\bibinfo {year} {2018})}\BibitemShut {NoStop}%
\bibitem [{\citenamefont {Ishimoto}\ and\ \citenamefont
  {Lauga}(2019)}]{ishimoto2019}%
  \BibitemOpen
  \bibfield  {author} {\bibinfo {author} {\bibfnamefont {K.}~\bibnamefont
  {Ishimoto}}\ and\ \bibinfo {author} {\bibfnamefont {E.}~\bibnamefont
  {Lauga}},\ }\bibfield  {title} {\enquote {\bibinfo {title} {The {N}-flagella
  problem: elastohydrodynamic motility transition of multi-flagellated
  bacteria},}\ }\href@noop {} {\bibfield  {journal} {\bibinfo  {journal} {Proc.
  R. Soc. A}\ }\textbf {\bibinfo {volume} {475}},\ \bibinfo {pages} {20180690}
  (\bibinfo {year} {2019})}\BibitemShut {NoStop}%
\bibitem [{\citenamefont {Belovs}\ and\ \citenamefont
  {C{\=e}bers}(2016)}]{belovs2016}%
  \BibitemOpen
  \bibfield  {author} {\bibinfo {author} {\bibfnamefont {M.}~\bibnamefont
  {Belovs}}\ and\ \bibinfo {author} {\bibfnamefont {A.}~\bibnamefont
  {C{\=e}bers}},\ }\bibfield  {title} {\enquote {\bibinfo {title}
  {Hydrodynamics with spin in bacterial suspensions},}\ }\href@noop {}
  {\bibfield  {journal} {\bibinfo  {journal} {Phys. Rev. E}\ }\textbf {\bibinfo
  {volume} {93}},\ \bibinfo {pages} {062404} (\bibinfo {year}
  {2016})}\BibitemShut {NoStop}%
\bibitem [{\citenamefont {Darnton}\ \emph {et~al.}(2007)\citenamefont
  {Darnton}, \citenamefont {Turner}, \citenamefont {Rojevsky},\ and\
  \citenamefont {Berg}}]{darnton2007}%
  \BibitemOpen
  \bibfield  {author} {\bibinfo {author} {\bibfnamefont {N.~C.}\ \bibnamefont
  {Darnton}}, \bibinfo {author} {\bibfnamefont {L.}~\bibnamefont {Turner}},
  \bibinfo {author} {\bibfnamefont {S.}~\bibnamefont {Rojevsky}}, \ and\
  \bibinfo {author} {\bibfnamefont {H.~C.}\ \bibnamefont {Berg}},\ }\bibfield
  {title} {\enquote {\bibinfo {title} {On torque and tumbling in swimming
  \textit{Escherichia coli}},}\ }\href@noop {} {\bibfield  {journal} {\bibinfo
  {journal} {J. Bacteriol.}\ }\textbf {\bibinfo {volume} {189}},\ \bibinfo
  {pages} {1756--1764} (\bibinfo {year} {2007})}\BibitemShut {NoStop}%
\bibitem [{\citenamefont {Hyon}\ \emph {et~al.}(2012)\citenamefont {Hyon},
  \citenamefont {Powers}, \citenamefont {Stocker},\ and\ \citenamefont
  {Fu}}]{hyon2012}%
  \BibitemOpen
  \bibfield  {author} {\bibinfo {author} {\bibfnamefont {Y.}~\bibnamefont
  {Hyon}}, \bibinfo {author} {\bibfnamefont {T.~R.}\ \bibnamefont {Powers}},
  \bibinfo {author} {\bibfnamefont {R.}~\bibnamefont {Stocker}}, \ and\
  \bibinfo {author} {\bibfnamefont {H.~C.}\ \bibnamefont {Fu}},\ }\bibfield
  {title} {\enquote {\bibinfo {title} {The wiggling trajectories of
  bacteria},}\ }\href@noop {} {\bibfield  {journal} {\bibinfo  {journal} {J.
  Fluid Mech.}\ }\textbf {\bibinfo {volume} {705}},\ \bibinfo {pages} {58--76}
  (\bibinfo {year} {2012})}\BibitemShut {NoStop}%
\bibitem [{\citenamefont {Bianchi}\ \emph {et~al.}(2017)\citenamefont
  {Bianchi}, \citenamefont {Saglimbeni},\ and\ \citenamefont
  {Di~Leonardo}}]{bianchi2017}%
  \BibitemOpen
  \bibfield  {author} {\bibinfo {author} {\bibfnamefont {S.}~\bibnamefont
  {Bianchi}}, \bibinfo {author} {\bibfnamefont {F.}~\bibnamefont {Saglimbeni}},
  \ and\ \bibinfo {author} {\bibfnamefont {R.}~\bibnamefont {Di~Leonardo}},\
  }\bibfield  {title} {\enquote {\bibinfo {title} {Holographic imaging reveals
  the mechanism of wall entrapment in swimming bacteria},}\ }\href@noop {}
  {\bibfield  {journal} {\bibinfo  {journal} {Phys. Rev. X}\ }\textbf {\bibinfo
  {volume} {7}},\ \bibinfo {pages} {011010} (\bibinfo {year}
  {2017})}\BibitemShut {NoStop}%
\bibitem [{\citenamefont {Drescher}\ \emph {et~al.}(2011)\citenamefont
  {Drescher}, \citenamefont {Dunkel}, \citenamefont {Cisneros}, \citenamefont
  {Ganguly},\ and\ \citenamefont {Goldstein}}]{drescher2011}%
  \BibitemOpen
  \bibfield  {author} {\bibinfo {author} {\bibfnamefont {K.}~\bibnamefont
  {Drescher}}, \bibinfo {author} {\bibfnamefont {J.}~\bibnamefont {Dunkel}},
  \bibinfo {author} {\bibfnamefont {L.~H.}\ \bibnamefont {Cisneros}}, \bibinfo
  {author} {\bibfnamefont {S.}~\bibnamefont {Ganguly}}, \ and\ \bibinfo
  {author} {\bibfnamefont {R.~E.}\ \bibnamefont {Goldstein}},\ }\bibfield
  {title} {\enquote {\bibinfo {title} {Fluid dynamics and noise in bacterial
  cell--cell and cell--surface scattering},}\ }\href@noop {} {\bibfield
  {journal} {\bibinfo  {journal} {Proc. Natl. Acad. Sci.}\ }\textbf {\bibinfo
  {volume} {108}},\ \bibinfo {pages} {10940--10945} (\bibinfo {year}
  {2011})}\BibitemShut {NoStop}%
\bibitem [{\citenamefont {Dominick}\ and\ \citenamefont
  {Wu}(2018)}]{dominick2018}%
  \BibitemOpen
  \bibfield  {author} {\bibinfo {author} {\bibfnamefont {C.~N.}\ \bibnamefont
  {Dominick}}\ and\ \bibinfo {author} {\bibfnamefont {X.-L.}\ \bibnamefont
  {Wu}},\ }\bibfield  {title} {\enquote {\bibinfo {title} {Rotating bacteria on
  solid surfaces without tethering},}\ }\href@noop {} {\bibfield  {journal}
  {\bibinfo  {journal} {Biophys. J.}\ }\textbf {\bibinfo {volume} {115}},\
  \bibinfo {pages} {588--594} (\bibinfo {year} {2018})}\BibitemShut {NoStop}%
\bibitem [{\citenamefont {Chen}\ \emph {et~al.}(2015)\citenamefont {Chen},
  \citenamefont {Yang}, \citenamefont {Yang},\ and\ \citenamefont
  {Zhang}}]{chen2015}%
  \BibitemOpen
  \bibfield  {author} {\bibinfo {author} {\bibfnamefont {X.}~\bibnamefont
  {Chen}}, \bibinfo {author} {\bibfnamefont {X.}~\bibnamefont {Yang}}, \bibinfo
  {author} {\bibfnamefont {M.}~\bibnamefont {Yang}}, \ and\ \bibinfo {author}
  {\bibfnamefont {H.~P.}\ \bibnamefont {Zhang}},\ }\bibfield  {title} {\enquote
  {\bibinfo {title} {Dynamic clustering in suspension of motile bacteria},}\
  }\href@noop {} {\bibfield  {journal} {\bibinfo  {journal} {Europhys. Lett.}\
  }\textbf {\bibinfo {volume} {111}},\ \bibinfo {pages} {54002} (\bibinfo
  {year} {2015})}\BibitemShut {NoStop}%
\bibitem [{ish()}]{ishimoto2019b}%
  \BibitemOpen
  \href@noop {} {}\bibinfo {note} {K.~Ishimoto, Bacterial spinning top, J.
  Fluid Mech. (to appear)}\BibitemShut {NoStop}%
\bibitem [{\citenamefont {Tornberg}\ and\ \citenamefont
  {Shelley}(2004)}]{tornberg2004}%
  \BibitemOpen
  \bibfield  {author} {\bibinfo {author} {\bibfnamefont {A.-K.}\ \bibnamefont
  {Tornberg}}\ and\ \bibinfo {author} {\bibfnamefont {M.~J.}\ \bibnamefont
  {Shelley}},\ }\bibfield  {title} {\enquote {\bibinfo {title} {Simulating the
  dynamics and interactions of flexible fibers in {S}tokes flows},}\
  }\href@noop {} {\bibfield  {journal} {\bibinfo  {journal} {J. Comp. Phys.}\
  }\textbf {\bibinfo {volume} {196}},\ \bibinfo {pages} {8--40} (\bibinfo
  {year} {2004})}\BibitemShut {NoStop}%
\bibitem [{\citenamefont {Pozrikidis}(2002)}]{pozrikidis2002}%
  \BibitemOpen
  \bibfield  {author} {\bibinfo {author} {\bibfnamefont {C.}~\bibnamefont
  {Pozrikidis}},\ }\href@noop {} {\emph {\bibinfo {title} {A Practical Guide to
  Boundary Element Methods with the Software Library BEMLIB}}}\ (\bibinfo
  {publisher} {Chapman \& Hall/CRC},\ \bibinfo {address} {Boca Raton},\
  \bibinfo {year} {2002})\BibitemShut {NoStop}%
\bibitem [{\citenamefont {Jeffery}(1915)}]{jeffery1915}%
  \BibitemOpen
  \bibfield  {author} {\bibinfo {author} {\bibfnamefont {G.~B.}\ \bibnamefont
  {Jeffery}},\ }\bibfield  {title} {\enquote {\bibinfo {title} {On the steady
  rotation of a solid of revolution in a viscous fluid},}\ }\href@noop {}
  {\bibfield  {journal} {\bibinfo  {journal} {Proc. Lond. Math. Soc.}\ }\textbf
  {\bibinfo {volume} {2}},\ \bibinfo {pages} {327--338} (\bibinfo {year}
  {1915})}\BibitemShut {NoStop}%
\bibitem [{\citenamefont {Stimson}\ and\ \citenamefont
  {Jeffery}(1926)}]{stimson1926}%
  \BibitemOpen
  \bibfield  {author} {\bibinfo {author} {\bibfnamefont {M.}~\bibnamefont
  {Stimson}}\ and\ \bibinfo {author} {\bibfnamefont {G.~B.r}\ \bibnamefont
  {Jeffery}},\ }\bibfield  {title} {\enquote {\bibinfo {title} {The motion of
  two spheres in a viscous fluid},}\ }\href@noop {} {\bibfield  {journal}
  {\bibinfo  {journal} {Proc. Roy. Soc. A}\ }\textbf {\bibinfo {volume}
  {111}},\ \bibinfo {pages} {110--116} (\bibinfo {year} {1926})}\BibitemShut
  {NoStop}%
\bibitem [{\citenamefont {Brenner}(1961)}]{brenner1961}%
  \BibitemOpen
  \bibfield  {author} {\bibinfo {author} {\bibfnamefont {H.}~\bibnamefont
  {Brenner}},\ }\bibfield  {title} {\enquote {\bibinfo {title} {The slow motion
  of a sphere through a viscous fluid towards a plane surface},}\ }\href@noop
  {} {\bibfield  {journal} {\bibinfo  {journal} {Chem. Engng. Sci.}\ }\textbf
  {\bibinfo {volume} {16}},\ \bibinfo {pages} {242--251} (\bibinfo {year}
  {1961})}\BibitemShut {NoStop}%
\bibitem [{\citenamefont {Maude}(1961)}]{maude1961}%
  \BibitemOpen
  \bibfield  {author} {\bibinfo {author} {\bibfnamefont {A.~D.}\ \bibnamefont
  {Maude}},\ }\bibfield  {title} {\enquote {\bibinfo {title} {End effects in a
  falling-sphere viscometer},}\ }\href@noop {} {\bibfield  {journal} {\bibinfo
  {journal} {Brit. J. Appl. Phys.}\ }\textbf {\bibinfo {volume} {12}},\
  \bibinfo {pages} {293} (\bibinfo {year} {1961})}\BibitemShut {NoStop}%
\bibitem [{\citenamefont {Dean}\ and\ \citenamefont
  {O'Neill}(1963)}]{dean1963}%
  \BibitemOpen
  \bibfield  {author} {\bibinfo {author} {\bibfnamefont {W.~R.}\ \bibnamefont
  {Dean}}\ and\ \bibinfo {author} {\bibfnamefont {M.~E.}\ \bibnamefont
  {O'Neill}},\ }\bibfield  {title} {\enquote {\bibinfo {title} {A slow motion
  of viscous liquid caused by the rotation of a solid sphere},}\ }\href@noop {}
  {\bibfield  {journal} {\bibinfo  {journal} {Mathematika}\ }\textbf {\bibinfo
  {volume} {10}},\ \bibinfo {pages} {13--24} (\bibinfo {year}
  {1963})}\BibitemShut {NoStop}%
\bibitem [{\citenamefont {O'Neill}(1964)}]{oneill1964}%
  \BibitemOpen
  \bibfield  {author} {\bibinfo {author} {\bibfnamefont {M.~E.}\ \bibnamefont
  {O'Neill}},\ }\bibfield  {title} {\enquote {\bibinfo {title} {A slow motion
  of viscous liquid caused by a slowly moving solid sphere},}\ }\href@noop {}
  {\bibfield  {journal} {\bibinfo  {journal} {Mathematika}\ }\textbf {\bibinfo
  {volume} {11}},\ \bibinfo {pages} {67--74} (\bibinfo {year}
  {1964})}\BibitemShut {NoStop}%
\bibitem [{\citenamefont {Loewenberg}\ and\ \citenamefont
  {Hinch}(1996)}]{loewenberg1996}%
  \BibitemOpen
  \bibfield  {author} {\bibinfo {author} {\bibfnamefont {M.}~\bibnamefont
  {Loewenberg}}\ and\ \bibinfo {author} {\bibfnamefont {E.~J.}\ \bibnamefont
  {Hinch}},\ }\bibfield  {title} {\enquote {\bibinfo {title} {Numerical
  simulation of a concentrated emulsion in shear flow},}\ }\href@noop {}
  {\bibfield  {journal} {\bibinfo  {journal} {J. Fluid Mech.}\ }\textbf
  {\bibinfo {volume} {321}},\ \bibinfo {pages} {395--419} (\bibinfo {year}
  {1996})}\BibitemShut {NoStop}%
\bibitem [{\citenamefont {Goldman}\ \emph {et~al.}(1967)\citenamefont
  {Goldman}, \citenamefont {Cox},\ and\ \citenamefont {Brenner}}]{goldman1967}%
  \BibitemOpen
  \bibfield  {author} {\bibinfo {author} {\bibfnamefont {A.~J.}\ \bibnamefont
  {Goldman}}, \bibinfo {author} {\bibfnamefont {R.~G.}\ \bibnamefont {Cox}}, \
  and\ \bibinfo {author} {\bibfnamefont {H.}~\bibnamefont {Brenner}},\
  }\bibfield  {title} {\enquote {\bibinfo {title} {Slow viscous motion of a
  sphere parallel to a plane wall--{I} {M}otion through a quiescent fluid},}\
  }\href@noop {} {\bibfield  {journal} {\bibinfo  {journal} {Chem. Engng.
  Sci.}\ }\textbf {\bibinfo {volume} {22}},\ \bibinfo {pages} {637--651}
  (\bibinfo {year} {1967})}\BibitemShut {NoStop}%
\bibitem [{\citenamefont {Chwang}\ and\ \citenamefont {Wu}(1975)}]{chwang1975}%
  \BibitemOpen
  \bibfield  {author} {\bibinfo {author} {\bibfnamefont {A.~T.}\ \bibnamefont
  {Chwang}}\ and\ \bibinfo {author} {\bibfnamefont {T.~Y.}\ \bibnamefont
  {Wu}},\ }\bibfield  {title} {\enquote {\bibinfo {title} {Hydromechanics of
  low-{R}eynolds-number flow. {P}art 2. {S}ingularity method for {S}tokes
  flows},}\ }\href@noop {} {\bibfield  {journal} {\bibinfo  {journal} {J. Fluid
  Mech.}\ }\textbf {\bibinfo {volume} {67}},\ \bibinfo {pages} {787--815}
  (\bibinfo {year} {1975})}\BibitemShut {NoStop}%
\end{thebibliography}%

\end{document}